\documentclass{ar-1col}

\usepackage[comma]{natbib}
\usepackage{url}

\newcommand{\msun}{\mbox{$M_{\odot}$}}
\newcommand{\Msun}{\mbox{$M_{\odot}$}}
\newcommand{\lsun}{\mbox{$L_{\odot}$}}

\newcommand{\logL}{\mbox{$\log (L/L_{\odot}$)}} 
\newcommand{\rsun}{\mbox{$R_{\odot}$}}

\newcommand{\Zsun}{\mbox{$Z_{\odot}$}}
\newcommand{\zsun}{\mbox{$Z_{\odot}$}}
\newcommand{\teff}{\mbox{$T_{\rm eff}$}}
\newcommand{\Teff}{\mbox{$T_{\rm eff}$}}

\newcommand{\mdot}{\mbox{$\dot{M}$}}
\newcommand{\Mdot}{\mbox{$\dot{M}$}}

\newcommand{\msunyr}{\mbox{$M_{\odot} {\rm yr}^{-1}$}}
\newcommand{\modf}{\mbox{$\dot{M} v_{\infty} R_*^{0.5}$}}

\newcommand{\ha}{H$\alpha$}
\newcommand{\PV}{P {\sc v}}
\newcommand{\apj}{{\it ApJ}}
\newcommand{\aap}{{\it A\&A}}
\newcommand{\mnras}{{\it MNRAS}}
\newcommand{\araa}{{\it ARA\&A}}

\jname{Xxxx. Xxx. Xxx. Xxx.}
\jvol{AA}
\jyear{YYYY}
\doi{10.1146/((please add article doi))}

\begin{document}

\markboth{Jorick S. Vink}{Hot Star Mass Loss}

\title{Theory and Diagnostics of Hot Star Mass Loss}

\author{Jorick S. Vink,$^1$ 
\affil{$^1$Armagh Observatory and Planetarium, Armagh, Northern Ireland, BT61 9DG; email: jorick.vink@armagh.ac.uk}}

\begin{abstract}
Massive stars have strong stellar winds that direct their evolution through the upper Hertzsprung-Russell diagram and determine the black hole mass function. Secondly, wind strength dictates the atmospheric structure that sets the ionising flux. Thirdly, the wind directly intervenes with the stellar envelope structure, which is decisive for both single star and binary evolution, affecting predictions for gravitational wave events.  
Key findings of current hot-star research include:

\begin{itemize}
\item{The traditional line-driven wind theory is being updated with Monte\\
Carlo and co-moving frame computations, revealing a rich {\it multi}-variate\\
behaviour of the mass-loss rate $\dot{M}$ in terms of $M$, $L$, Eddington $\Gamma$, $T_{\rm eff}$,\\ 
and chemical composition $Z$. Concerning the latter, $\dot{M}$ is shown to\\depend
on the iron (Fe) opacity, making Wolf-Rayet populations, and\\ gravitational wave events
dependent on {\it host} galaxy $Z$.}

\item{On top of smooth mass-loss behaviour, there are several {\it transitions}\\ 
in the Hertzsprung-Russell diagram, involving\ bi-stability jumps around\\ Fe 
recombination temperatures, leading to quasi-stationary {\it episodic},\\
and not necessarily eruptive, Luminous Blue Variable and pre-SN\\
mass loss.}
\item{Moreover, there are {\it kinks}. At 100\,$\msun$ a high $\Gamma$ mass-loss transition\\  
implies that hydrogen-rich very massive stars have {\it higher}
mass-loss \\
rates than commonly considered. At the other end of the mass\\ spectrum, low-mass stripped helium stars no longer appear as \\ Wolf-Rayet stars, but
as optically-thin stars. These stripped stars, in \\addition 
to very massive stars, are two newly identified sources of \\ionising radiation that
could play a key role in local star formation as \\well as at high-redshift.}
\end{itemize}
\end{abstract}

\begin{keywords}
stellar winds, mass loss, stellar evolution, massive stars, clumps, supernovae
\end{keywords}
\maketitle

\tableofcontents

\section{Introduction}

Accurate mass-loss rates $\dot{M}$ for hot massive stars are critical for our understanding of the evolution and fate of the most massive stars ($M \ge 25\,\Msun$) in the Universe.
While the stellar initial mass function (IMF)
dictates that there are more massive stars below 25\,\Msun\ than above it, due to their higher luminosities $L_{\ast}$ and mass-loss rates
it is these objects more massive than 25\,\Msun\ that not only dominate the radiative and mechanical feedback, but it is also their
light that is expected to be detected with new facilities such as the James Webb Space Telescope (JWST) and ground-based extremely large telescopes (ELTs) in the coming decade.

\begin{marginnote}[]
\entry{Mass-loss rate}{follows from the mass-continuity equation: $\dot{M} = 4 \pi r^2 v(r) \rho(r)$ and is usually given in $\msunyr$.}
\end{marginnote}

\begin{marginnote}[]
\entry{Massive stars}{Stars above 8\,\Msun\ that make core-collapse SNe. Stars above 25\,\Msun\ may not produce SNe, but directly collapse into a BH instead. It is these stars above approx 25\,\Msun\ where winds are sufficiently strong to affect stellar evolution, which this Review is focused on.}
\end{marginnote}

Furthermore, the interest in the most massive stars above
25\,\Msun\ has grown tremendously thanks to the discovery of heavy black holes (BHs) of over 25\,\Msun\ via gravitational waves with LIGO/Virgo
(Abbott et al. 2016; Abbott et al. 2020). More generally, while there is an established paradigm for the evolution of massive OB
star in the 8-25\,\Msun\ range towards red supergiants (RSGs) leading to hydrogen (H) rich type IIP supernovae (SNe) (Smartt 2009), that likely leave
behind a neutron star, the evolutionary route for stars above 25\,\Msun\ is -- despite significant theoretical and observational progress -- still in its infancy (Langer 2012; Yusof et al 2013; Woosley \& Heger 2015; Limongi \& Chieffi 2018; Spera et al. 2019). Answers regarding the evolution of the most massive stars will rely on theoretical and observational progress in our detailed understanding of stellar winds -- as a function of metallicity $Z$: $\dot{M} = f(Z)$. 
Also the landscape of various SN sub-types (see SN Taxonomy) is thought to involve a mass-loss sequence, with the common IIP SNe still having the largest H envelope intact, while progressively higher mass stars loose more of their envelopes. 
This mass loss could involve $Z$-dependent winds, eruptive mass loss that might be $Z$-independent (Smith 2014), or alternatively due to Roche-Lobe overflow (RLOF) in a binary system. Mapping SN subtypes as a function of host galaxy $Z$ may reveal the relative importance of $Z$-dependent vs. $Z$-{\it in}dependent mass-loss mechanisms.

\begin{textbox}\section{Taxonomy for core collapse SNe}

H-rich Types II: 
\begin{itemize}
\item{IIP:~~~Plateau: Photometric long Plateau due to large H envelope.}
\item{IIL:~~~Linear: Photometric linear Plateau due to diminished H envelope.}
\item{IIn:~~~narrow: Spectroscopic narrow lines due to CSM interaction.}
\item{IIb:~~~transitional: very little H.}
\end{itemize}

H-poor Types I:
\begin{itemize}
\item{Ib:~~~No H.}
\item{Ic:~~~No H and No He in the spectrum.\\}
\end{itemize}

\begin{itemize}
\item{SLSNe:~ superluminous (Types I and II) are 2 orders of magnitude brighter.}
\item{PISNe:~ hypothezied Pair-Instability SNe. }
\end{itemize}
\end{textbox}

\begin{marginnote}[]
\entry{HD-limit}{Observed lack of RSGs above \logL\ = 5.8. Lower temperatures (Levesque et al. 2005) and luminosities \logL\ = 5.5 (Davies et al. 2018) nowadays place it at around 25\,\msun.}
\end{marginnote}

Stars up to approx 25\,$M_{\odot}$ become RSGs keeping their H envelopes intact. The evolutionary paths for more massive stars must be notably different as there is a complete absence of RSGs in the HR diagram above a critical luminosity limit: 
the Humphreys-Davidson (HD) limit
at a luminosity of \logL\ = 5.8 (Fig.\,1). This 
{\it bifurcation} between the H-rich and H-poor Universe 
forms a distinct feature in the Hertzsprung-Russell diagram (HRD).
While it has become clear that somewhere around 50\% (Sana et al. 2013; Kobulnicky 2014)\footnote{Recent imaging studies suggest that the {\it total} binary fraction is 
  close to 100\% (e.g. Sana et al. 2014), but some companions are too distant to alter massive-star evolution. It is the close binaries of approx 50\% discovered through radial velocity (RV) variations that are relevant for stellar evolution, involving the stripping of the H envelope, and even merging events on and off the main sequence (e.g. Eldridge \& Stanway 2022)} of massive OB stars are part of a close binary system, there is no a priori
reason that this would lead to such a sharp feature in the HRD. Instead, this tipping point is most likely related to the Eddington limit against radiation pressure (Humphreys \& Davidson 1979).

\begin{marginnote}[]
\entry{Eddington Limit}{Balance between radiative acceleration and inwards directed gravitational acceleration.}
\end{marginnote}

\begin{figure*}
\begin{centering}
\includegraphics[width=\textwidth]{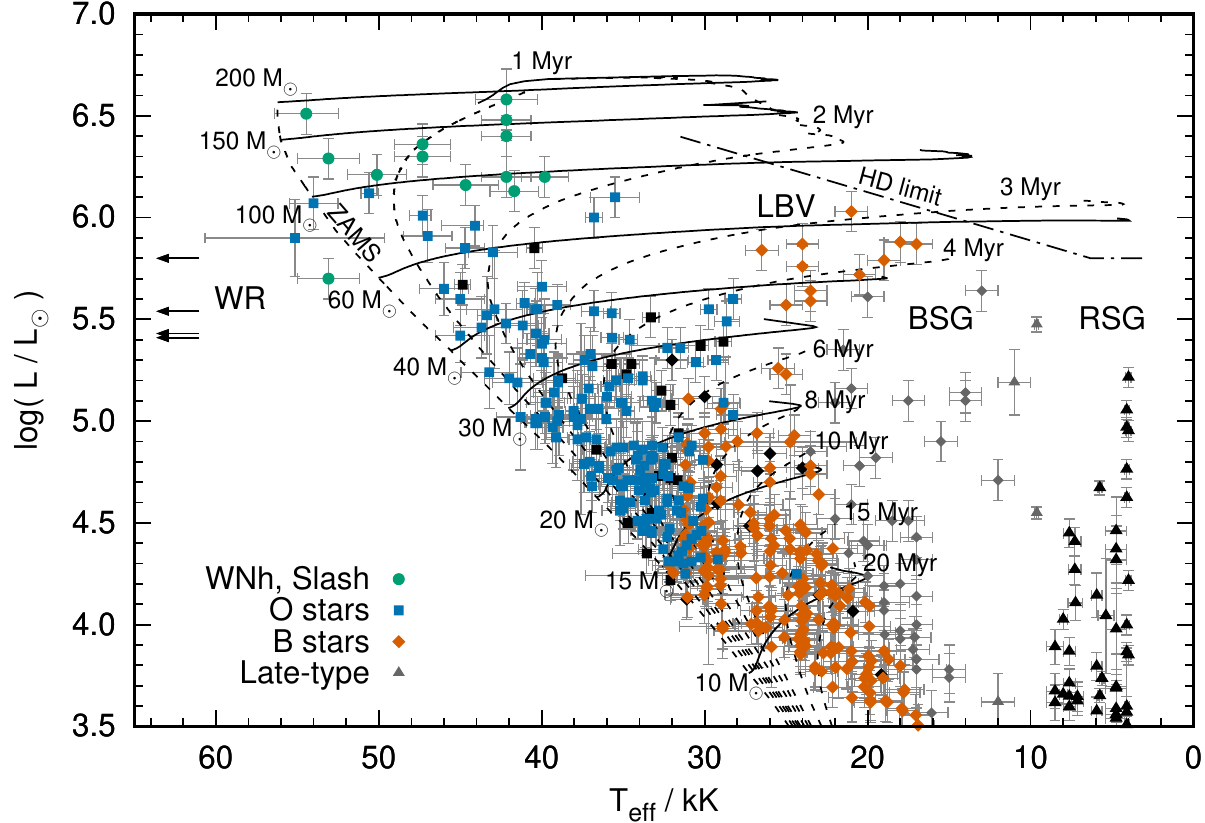}
\par\end{centering}
\caption{Hertzsprung--Russell diagram of the VLT-Flames Tarantula Survey (Evans et al. 2011) in the LMC. 
The solid and dashed black lines are the non-rotating Bonn stellar evolutionary tracks and isochrones from Brott et al. (2011) and K\"ohler et al. (2015). 
The original HD-limit is shown by the dot-dashed line in the top-right corner. 
Symbols are indicated in the legend. 
Slash stars are Of/WN stars.
Objects cooler than 9000\,K are shown as triangles. 
The luminosities of the 
classical WR stars are indicated by the arrows on the left. They actually have surface temperatures of over 90\,kK. Figure adapted from Schneider et al. (2018).}
\label{fig_hrd}
\end{figure*}

In any case -- contrary to the lower mass range -- stars above 25\msun\ do not become RSGs.
This implies that it is fast {\it hot-star winds} with $\teff \ge 10,000$K that dominate the mass-loss history and fate of the most massive stars that need to be quantitatively mapped
to understand H-rich and H-poor (super-luminous) SNe, and to accurately predict the BH mass function as a function of metallicity $Z$.
While during the last 50 years a theoretical paradigm for these hot-star winds in terms of radiation pressure has been established (Castor et al. 1975; hereafter CAK),
there are still many uncertainties in even the absolute mass-loss rates
predicted by the radiation-driven wind theory (RDWT). In the 8-25\,\Msun\ regime observational results have been published where the discrepancies between the RDWT of Vink et al. (2000)
and observations have been
up to 2 orders of magnitude in the {\it weak-wind regime} (Martins et al. 2005).

\begin{marginnote}[]
\entry{VMS}{Stars more massive than 100\msun. In the local Universe these are WNh stars: N-type WR stars that still contain H. The less dense-winded Of/WN "slash" stars are usually also included.}
\end{marginnote}

By contrast, on the other extreme, for very massive stars (VMS) of order 100\,\Msun\ and above (Crowther et al. 2010; Martins 2015),
we know the absolute mass-loss rate relatively accurately (within approx. 30\%; Vink \& Gr\"afener 2012), but because the cumulative effect of mass loss on stellar evolution models
is more dominant in this regime, the uncertainties in mass-loss history are nonetheless profound. This is especially true given that 
VMS are close to the Eddington limit (Hirschi 2015; Owocki 2015) 
and there is not yet any settled solution for stellar structure models as to how the envelopes and stellar winds interact with one another. This implies that VMS radii and effective temperatures are highly uncertain, and we thus need to unravel how the mass-loss rates depends on \Teff. This is absolutely critical for establishing whether VMS are important sources of ionising radiation in the Universe.

Despite the fact that mass-loss rates for approx 100\,\Msun\ stars have been accurately calibrated at the transition mass-loss rate (see Sect.\,\ref{sec_trans}), for stars in the canonical O star regime above 25\,\Msun\ there are
uncertainties in mass-loss rates of a factor of 2-3 (e.g. Bouret et al. 2013; Smith 2014; Bj\"orklund et al. 2021).
These uncertainties are mostly related to wind clumping, the radiative transfer treatment in
sub-sonic regions, and wind hydrodynamics beyond CAK. Finally, note that an uncertainty of just a factor 2 in the mass-loss rate in
stellar evolution of a canonical O star has long been known to be crucial (e.g. Meynet et al. 1994).
Such a relatively small difference in $\mdot$ on the main sequence can make a 60\,\Msun\ star either remain intact and die as a $\sim$25\,\Msun\ BH, or almost completely "evaporate" dying as a neutron star or "light BH" of just a few solar masses.

In other words, it is pertinent that we develop an accurate framework for radiation-driven winds over the entire upper HRD, and that we predict how the mass-loss rate depends on the stellar mass $M$, luminosity $L$
(or Eddington parameter $\Gamma$), the effective temperature \Teff, and chemical composition. Moreover, stars may not remain in the same HRD position, but change their \teff\ locations on relatively short timescales (of order years and decades for the case of luminous blue variables; LBVs). This could lead to episodic mass loss that could either be described by stationary wind models, or in other situations the
mass-loss rates are so humongous ($\sim 10^{-2} \msunyr$ or higher) that non-stationary processes such as eruptive mass loss or dynamical phenomena such as explosions need to be considered, especially  for
giant eruptions of LBVs (e.g. Shaviv 2000; Owocki 2015) and certain SN progenitors, such as the super-luminous IIn SN 2006gy (Smith et al. 2007).
Note that the existence of these extreme cases where super-Eddington winds and explosions are required, does {\it not} imply 
that each and every case of episodic or pre-SN mass loss
demands such physical extremes. 
It is quite possible that for the vast majority of pre-SN mass-loss cases (quasi)stationary winds can explain most episodic mass loss, and we will return to this in Sect.\,\ref{sec_lbv}.

For convenience I provide an overview of mass-loss recipes for hot-star winds in the Appendix. 
For a recent overview on cooler red supergiant (RSG) winds, I simply refer to Decin (2020). This review will focus on hot star winds from massive stars only, but general principles should apply to low-mass stars as well.

\section{Radiation-driven wind Theory}

\begin{marginnote}[]
\entry{P Cygni profile}{Characteristic spectral resonance line involving almost symmetric line emission and blue-shifted absorption. Named after the LBV P\,Cygni. See the right-hand side of Fig.\,\ref{f_oski} for an example.}
\end{marginnote}

\begin{marginnote}
\entry{Wolf-Rayet stars}{Spectroscopic designation based on emission lines. WRs come in many flavours but the {\it classical} WR stars are of types WN and WC: the nitrogen-rich WN stars have enhanced nitrogen (N) due to the CNO cycle, while the more evolved WC stars show the product of He burning -- in the form of carbon (C) -- at the surface.} 
\end{marginnote}

While it had already been established in the early parts of the 20th century that radiation pressure was capable of ejecting atoms from stellar atmospheres, the theory matured in the 1970s thanks to observational impetus from ultraviolet (UV) spectroscopy of O-type stars with satellites such as the International Ultraviolet Explorer (IUE). These data showed the widespread presence of P\,Cygni profiles in UV resonance lines, such as carbon {\sc iv}, silicon {\sc iv} and nitrogen {\sc v} which were interpreted as stationary outflows that might be sufficiently strong to affect massive star evolution (Morton 1967, Lamers \& Morton 1976). The first theoretical models predicted modest rates (Lucy \& Solomon 1970) but when CAK introduced line-distribution functions the predicted O-star rates grew by 2 orders of magnitude to approx $10^{-6} \msunyr$ -- sufficient to affect stellar evolution. This lead to the so-called "Conti" (1975) scenario where the loss of significant amounts of envelope mass would result in the formation of classical Wolf-Rayet (WR) stars. The mass-loss rates predicted by CAK theory are now thought to be somewhat on the high side (Smith 2014), but still correct within a factor of a few (Puls et al. 2008). 
In the 1980s, the "Munich Group" (e.g. Pauldrach et al. 1986) updated the atomic data with the inclusion of millions of iron (Fe) lines and established a mass-loss metallicity ($Z$) relation (Abbott 1982; Kudritzki et al. 1987). These models up to the year 2000 are referred to as modified CAK (or M-CAK) models, as the CAK theory was modified to include the effects of the finite-disk of the star with non-radially streaming photons. 

In the new millennium most stellar evolution models switched to using the Monte Carlo predictions of Vink et al. (2000) that were based on a technique developed by Abbott \& Lucy (1985). These models included the effect of multiple photon scattering in a realistic way, but the hydrodynamics was simplified in that the models relied on a global energy approach. M\"uller \& Vink (2008) improved this assumption, providing locally consistent models through the use of the Lambert W function (see also Muijres et al. 2012; Gormaz-Matamala 2021) but still relying on the Sobolev approximation -- a questionable assumption in the sub-sonic portions of the wind. 

\begin{marginnote}
\entry{Lambert W Function}{A function which allows one to solve mathematical problems where the variable is both a base and an exponent, i.e. $f(w) = w\exp{w}$, where $w$ is any complex number (Corless et al. 1993)}.
\end{marginnote}

\begin{marginnote}[]
\entry{Sobolev approximation}{Used to simplify radiative transfer when the velocity gradient $dv/dr$ is large. i.e. when conditions over the Sobolev length $v_{\rm th}/(dv/dr)$ -- in terms of the thermal line width ($v_{\rm th}$) -- hardly change.}
\end{marginnote}

This is why during the last 5 years progress has been made using non-LTE co-moving frame (CMF) approaches using codes such as {\sc cmfgen}, {\sc PoWR}, {\sc fastwind} and {\sc metuje} that no longer rely on the Sobolev approximation (Petrov et al. 2016; Sander et al. 2017; Bjorklund et al. 2021; Krticka \& Kubat 2017).
 
\begin{marginnote}[]
\entry{non-LTE}{It literally means that the gas is {\it not} in Local Thermal Equilibrium. But here it indicates that the statistical equilibrium (SE) equations are explicitly solved for, without relying on Saha-Boltzmann approximations.}
\end{marginnote}

\subsection{CAK Theory}

The basic idea of CAK theory is to describe the radiative force $a_{\rm rad}$ in terms of a force multiplier $M$. 
This is a way to describe the radiative force on the opacity resulting from Doppler-shifted lines through multiples of the electron scattering opacity. 
For hot O and WR stars H and He are fully ionised, and the electron scattering opacity consequently easily established. The challenge is to compute $M$ in terms of the Sobolev velocity gradient $(dv/dr)_{\rm Sob}$ (to the power $\alpha$, where the famous CAK-parameter $\alpha$ gives the line force from optically thin lines to the total line force, related to the line-strength distribution, e.g. Puls et al. 2000). 
By entering the radiative acceleration $a_{\rm rad}$ into the equation of motion, and neglecting the gas pressure, an excellent assumption in the supersonic part of the wind, one only needs to consider:

\begin{equation}
v~(\frac{dv}{dr})~=~-\frac{GM_{\ast}}{r^2}~+~C~(\frac{dv}{dr})^{\alpha}_{\rm Sob}
\label{eq_eom}
\end{equation}
Where $C$ is a constant. By equating the velocity gradient in the inertia term on the left-hand side of Eq.\,\ref{eq_eom} to the Sobolev velocity gradient on the right-hand side, CAK could 
solve the equation of motion analytically, and predict the wind velocity to behave as a "beta" law:

\begin{equation}
v(r)~=~v_{\infty}~(1 - \frac{R_{\ast}}{r})^{\beta} 
\end{equation}
\label{eq_vel}
with $\beta = 0.5$ and a mass-loss rate $\mdot$ that scales with both the stellar mass $M_{\ast}$ and Eddington factor $\Gamma$, or $L_{\ast}$. Through the 1980s and 1990s the CAK models were modified using more complex spectral line-lists. Furthermore, the point-source approximation was lifted resulting in a somewhat larger $\beta$ values of order 0-8\,-\,1, and with terminal wind velocities now 2-3 times the stellar escape velocity, numerically corresponding to values in the correct range of 2000-3000 km/s for O-type stars (e.g. Pauldrach et al. 1986). I refer to Puls et al. (2008) and Owocki (2015) for more detailed derivations and discussions regarding (modified) CAK theory, before ending this subsection with a clever "trick" of Kudrizki et al. (1995): they constructed the so-called modified wind momentum rate,
$D_{\rm mom}\,=\,\mdot\,v_{\infty}\,(R/\rsun)^{1/2}$. Given that $v_{\infty}$ scales
with the escape velocity, and as long as the CAK parameter $\alpha$ is constant and of order 2/3, the stellar mass conveniently cancels out (the "trick"), resulting in a momentum product that only scales with $L_{\ast}$: 

\begin{marginnote}
\entry{Point Source approximation}{The star's finite sized "disk" is assumed to be infinitely small so that the star is effectively a point source -- with photons streaming out radially.}
\end{marginnote}

\begin{equation}
\log D_{\rm mom}~=~x~\logL  + D_{0},
\label{wlr}
\end{equation}
with slope $x$ and offset $D_{0}$, the ``wind momentum
luminosity relationship (WLR)'' 
(Kudritzki et al. 1995) was born.
While initially proposed to be a cosmological distance indicator, the WLR played an instrumental role in determining the
empirical mass-loss metallicity dependence for O stars in 
the Small and Large Magellanic Clouds (Mokiem et al. 2007). Today, observed and predicted WLRs 
can be compared to test the validity of the theory in different physical regimes, helping to highlight potential 
shortcomings, e.g. regarding wind clumping. One should however be aware that 
in reality $\dot{M}$ is not only a function of $L_{\ast}$ but also of parameters such as mass $M_{\ast}$ and 
$T_{\rm eff}$, and one should properly account for this multi-variate behaviour of $\dot{M}$ 
when comparing observations to theory.
Finally, CAK-type relations are only valid 
for spatially constant CAK Force Multiplier parameters, an assumption that does not hold in more realistic models 
(Vink 2000; Kudritzki 2002; Muijres et al. 2012; Sander et al. 2020).

\subsection{Beyond CAK}

Despite the plethora of achievements by the (modified) CAK theory discussed above, it was clear that in order to achieve a more realistic framework of the radiation-driven wind theory (RDWT) a number of assumptions needed to be improved upon. 
The main drawbacks of the (modified) CAK theory are the usage of the Sobolev approximation (see Sect.\,\ref{sec_cmf}), the simplified hydrodynamics in terms of a 2-parameter Force Multiplier approach, the reliance on the CAK critical point, and the neglect of multiple scattering. 

\subsubsection{Multiple scattering and the Monte Carlo Method}

The first alternative approach to CAK to be discussed is 
the Monte Carlo (MC) method developed by 
Abbott \& Lucy (1985). In this approach photon packets are 
tracked on their journey from the photosphere to the outer wind. At each interaction, momentum and energy 
are transferred from the photons to the gas particles. 
One of the major advantages of the Monte Carlo method is that it includes multi-line scatterings in a relatively straightforward 
manner (see Schmutz et al. 1990; de Koter et al. 1997; Noebauer \& Sim 2019). 
Prior to the year 2000, (modified) CAK mass-loss rates fell short of the observed rates
for the denser O star and WR winds.
The crucial point is that multiple scattered photons add radially outward
momentum to the wind, and the momentum may exceed the 
single-scattering limit. Phrased differently, the wind efficiency 
$\eta = \mdot v_{\infty}/(L_{\ast}/c)$ can exceed unity (see Sect.\,\ref{sec_trans}). 

\begin{figure*}
\begin{centering}
\includegraphics[width=\textwidth]{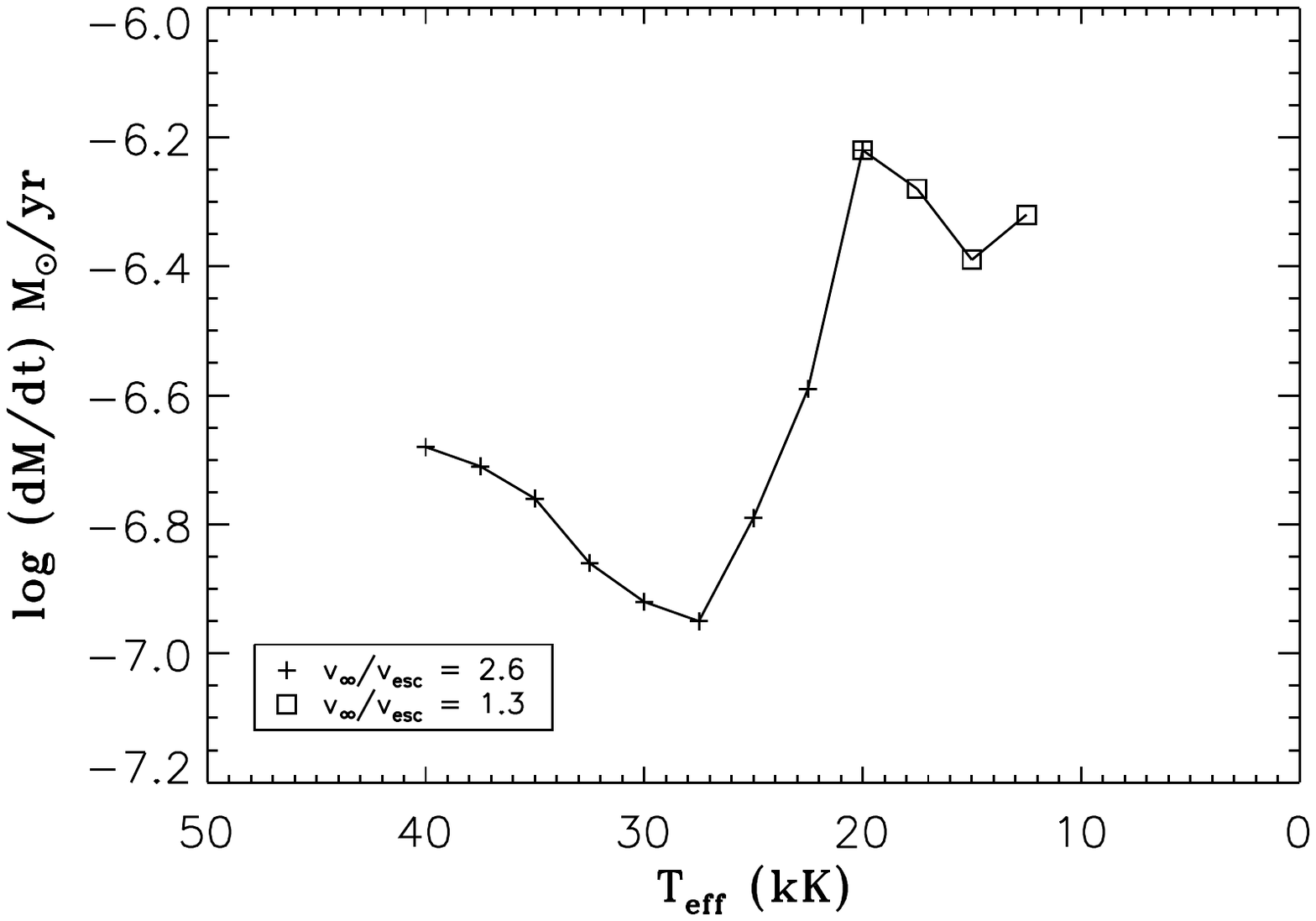}
\par\end{centering}
\caption{Monte Carlo mass-loss rate predictions as a function of \teff \ across the bi-stability regime, where $\dot{M}$ is predicted to jump upwards by a factor of approx five, even for this moderately massive star. In these older "global" models the ratio of $v_{\infty}/v_{\rm esc}$ was assumed to drop -- by a factor of two -- from 2.6 to 1.3. Adapted from Vink et al. (1999). }
\label{f_bsj}
\end{figure*}

\begin{marginnote}
\entry{Bi-stability Jump}{The BS-Jump forms a bi-furcation in wind physics at spectral type B1 (approx 21 kK) where on the hot side winds are fast and of modest strength, while they become slower and stronger on the cool side of the critical temperature. }
\end{marginnote}

\begin{marginnote}
\entry{Weak-wind regime}{An empirical regime where mass-loss rates below luminosities of $\log L/\lsun = 5.2$ are approx 2 orders of magnitude lower than predicted. This roughly coincides with absolute mass-loss rates below $10^{-7}$ \msunyr.}
\end{marginnote}

Vink et al. (1999, 2000) used the Monte Carlo method to derive 
a mass-loss recipe (see Appendix), where for objects hotter than the 
so-called bi-stability jump (BS-Jump; see Fig.\,\ref{f_bsj}) at $\simeq$ 20-25\,kK, the rates roughly scale as:
\begin{equation}
\mdot~\propto~L_{\ast}^{2.2}~M_{\ast}^{-1.3}~\Teff~(\frac{v_{\infty}}{v_{\rm esc}})^{-1.3}
\label{eq_formula}
\end{equation}
The main success of these models was that they were equally successful for relatively 
weak winds (with \mdot\ down to  $10^{-7}$ \msunyr)\footnote{This does not include the "weak wind regime" (Martins et al. 2005; Puls et al. 2008; de Almeida et al. 2019).} as dense O-star 
winds (with \mdot\ as high as $10^{-4}$ \msunyr). Note that the positive $\dot{M}$ dependence with \teff, translates into a dropping mass-loss rate with \teff\ during redwards stellar evolution. 
The reason is that while the flux distribution shifts to longer wavelengths, 
most line-transition opacities remain at shorter wavelengths -- resulting in a growing mismatch between the flux and the opacity.
This changes abruptly when Fe {\sc iv} recombines to Fe {\sc iii} in the stellar atmosphere. 
With the increased amount of line transitions in Fe {\sc iii}, the flux-weighted opacity increases, which defines the bi-stability jump.
Note that the dual dependence of $\dot{M}$ on both $M_{\ast}$ and $L_{\ast}$ in Eq.\,\ref{eq_formula} can in this theoretical setting be simplified to the somewhat shallower relation $\dot{M} \propto L_{\ast}^{1.6}$ using the Mass-Luminosity relation. One has to take care when stellar masses are not known, and one cannot simply replace an $L$-dependence that includes a $M_{\ast}$ dependence with another one that does not.  

The predictions of the mass-loss recipe are only valid for hot-star winds 
at a sufficient distance from the Eddington limit, with $\Gamma$
$<$ 0.5. There are two regimes where this is no longer the case: (i)
stars that have formed with high initial masses and luminosities, i.e. very
massive stars (VMS), and (ii) less extremely
luminous stars that approach the Eddington limit when they have
evolved significantly. Examples of this latter category are LBVs and classical 
WR stars. 
Vink \& de Koter (2002) showed that the
mass-loss rate for LBVs increase more rapidly with $L/M$ than Eq.~(\ref{eq_formula}) would indicate.
For VMS, Vink et al. (2011) discovered a kink (see Fig.\,\ref{f_mdotgamma}) in the
mass-loss vs. $\Gamma$ relation -- with the slope jumping from 2 to 5 -- right at the transition from optically thin O-type
to optically thick WNh winds (see Sect.\,\ref{sec_trans}). 

The main drawback of the MC method -- but simultaneously also its strength -- was the initial assumption of a pre-determined 
velocity structure. As the terminal wind velocity is an accurate empirically derived number ($\sim$10\% from the blue edge of UV P\,Cygni lines), the velocity structure in the supersonic wind part -- where the Sobolev approximation is excellent -- was basically enforced to be correct, and predicted mass-loss rates from the Monte Carlo method could thus be considered to be rather accurate. 
This also implied that they were useful as input for stellar evolution modelling.
However, from a purely theoretical hydrodynamical perspective, the global Monte Carlo approach was rather unsatisfactory, which is why the 
modelling was improved involving a locally consistent "Lambert W" framework by M\"uller \& Vink (2008) for 1D and M\"uller \& Vink (2014) for 2D.

Solving the equation of motion self-consistently, and no longer relying on free parameters, M\"uller \& Vink (2008) 
determined the velocity field through the use of the Lambert W function with a line acceleration 
that only depended on radius (rather than explicitly on the velocity gradient $dv/dr$ as in CAK theory.) 
As a result of this simplified $a_{\rm rad}(r)$ dependency, the CAK critical point was no longer the critical point of the wind 
(see Abbott 1982; Lucy 2010), but this had now become 
the sonic point (as in the Parker (1958) solar wind theory). 
For the iso-thermal case, M\"uller \& Vink (2008) derived an analytic solution of the velocity law 
which was compared to the $\beta$-law and utilised to derive $v_{\infty}$ values. 
These were approx. 25-40\% higher than observed, and due to the invariance of the momentum product $\dot{M}$ $v_{\infty}$, mass-loss rates somewhat
lower. Similar small over-predictions in terminal wind velocity occur in MCAK and recent CMF computations of Bj\"orklund et al. (2021), possibly highlighting that some physics is still missing in 1D models.
Muijres et al. (2012a) tested the M\"uller \& Vink (2008) 
wind solutions through explicit numerical integrations of the 
fluid equation, also accounting for a temperature stratification, obtaining 
results that were in good agreement with the M\"uller \& Vink solutions.

\subsubsection{Co-moving Frame method}
\label{sec_cmf}

Despite the many successes of the CAK and MC methods, one of the remaining drawbacks is the use of the Sobolev approximation. 
The Sobolev approximation is considered excellent in the supersonic parts of stellar outflows, but it falls short around and below the sonic point, where the mass-loss rate in is initiated. It is thus pertinent that a full CMF approach to the radiative transfer is contemplated. Over the last $\sim$5 years, 
a number of CMF calculations of the radiative transfer have been published. 
These are based on complex non-LTE atmospheric codes, such as CMFGEN (Hillier \& Miller 1998), PoWR (e.g. Gr\"afener et al. 2002), and {\sc fastwind} (Puls et al. 2020) that had traditionally only been used to 
simulate the spectra of hot massive stars for comparison with observations, with the aim of empirically determining the stellar parameters such as $\log\,g$ and $\teff$ as well as wind parameters such as $\mdot$. However, these empirical studies have until recently always been performed through the use of an adopted constant $\beta$ velocity law $v(r)$. While the task of constructing non-LTE model atmospheres would deserve a review in its own right, here I will just discuss the new hydrodynamical aspects of these sophisticated codes. I refer to the main publications of the codes for 
their specific physical inputs (such as the treatment of atomic data, the temperature stratification, etc.). 
For the art of spectral analysis employing these model atmosphere codes, see the recent review by Simon-Diaz (2020).   

Instead of parameterising the radiative acceleration, the current state-of-the art in terms of hydrodynamics relies on a direct integration of the radiative acceleration in the CMF, 

\begin{equation}
 \label{eq:arad}
  a_{\rm rad} = \frac{1}{c} \int \kappa_\nu F_\nu {\rm d}\nu,
\end{equation}
Where $\kappa_{\nu}$ is the frequency dependent opacity and $F_{\nu}$ the flux. This full integration has been performed by Gr\"afener \& Hamann (2005) using {\sc PoWR}, Bjorklund et al. (2021) using {\sc fastwind}, Krticka \& Kubat (2017) using {\sc metuje} and Gormaz-Matala (2021) using the Lambert W concept in {\sc cmfgen}\footnote{For an earlier global energy method with {\sc cmfgen} see Petrov et al. (2016)}. 
The advantage over previous approaches is that the Sobolev approximation is no longer used, and as long as the detailed atmospheric physics allows for it, 
these CMF calculations are equally applicable for optically thin as for optically thick winds, such as the winds of WR stars (Sander et al. 2020), although a monotonic velocity field is enforced.

\subsubsection{Optically thick winds}  

Classical WR stars have strong winds with large mass-loss rates (Crowther 2007), typically a factor of 
10 larger than O-star winds with the same luminosity, and they are not at all explicable by optically thin CAK theory involving a $dv/dr$ gradient. By contrast in these optically thick wind models $dv/dr \simeq 0$. 
The observed wind efficiency $\eta$ values are
typically in the range of 1-5, i.e. well above the single-scattering limit (e.g. Hainich et al. 2014; Hamann et al. 2019).
As the state of ionisation decreases outwards, photons may interact with opacity from a variety of different 
ions (such as iron, Fe) on their way out, whilst gaps 
between lines become ``filled in'' due to the layered ionisation structure (Lucy \& Abbott 1993; Gayley et al. 1995). 
The initiation of classical WR outflows relies on
the condition that the winds are optically thick at the sonic point, and that the line acceleration due
to the high opacity ``iron peak'' is able to overcome gravity, driving an optically thick WR wind (Nugis \& Lamers 2002). 

\begin{figure}
  \includegraphics[width=\columnwidth]{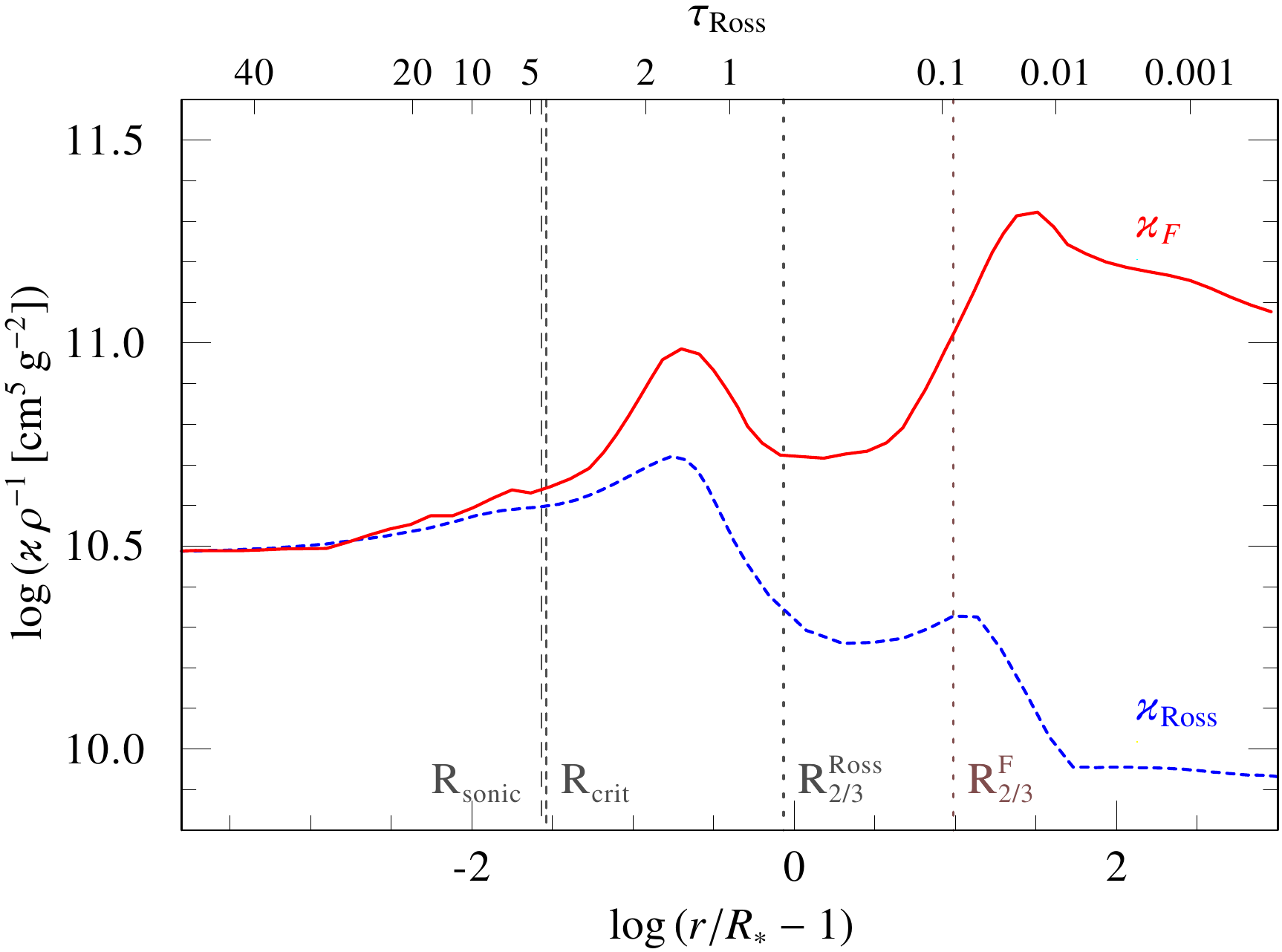}
  \caption{Comparison of the flux-weighted mean opacity (red, solid) to the Rosseland mean opacity (blue, dashed) in a model atmosphere model for a hydrogen-free N-type WR star (see Sander et al. (2020) for parameters) at solar metallicity. Note that the Rosseland and flux-mean opacities are almost equal each other at the sonic (and critical) point, but that they diverge by an order of magnitude in the outer wind. The 2 iron bumps are notably present.}
  \label{f_checkopaross}
\end{figure}

\begin{figure}
  \includegraphics[width=\columnwidth]{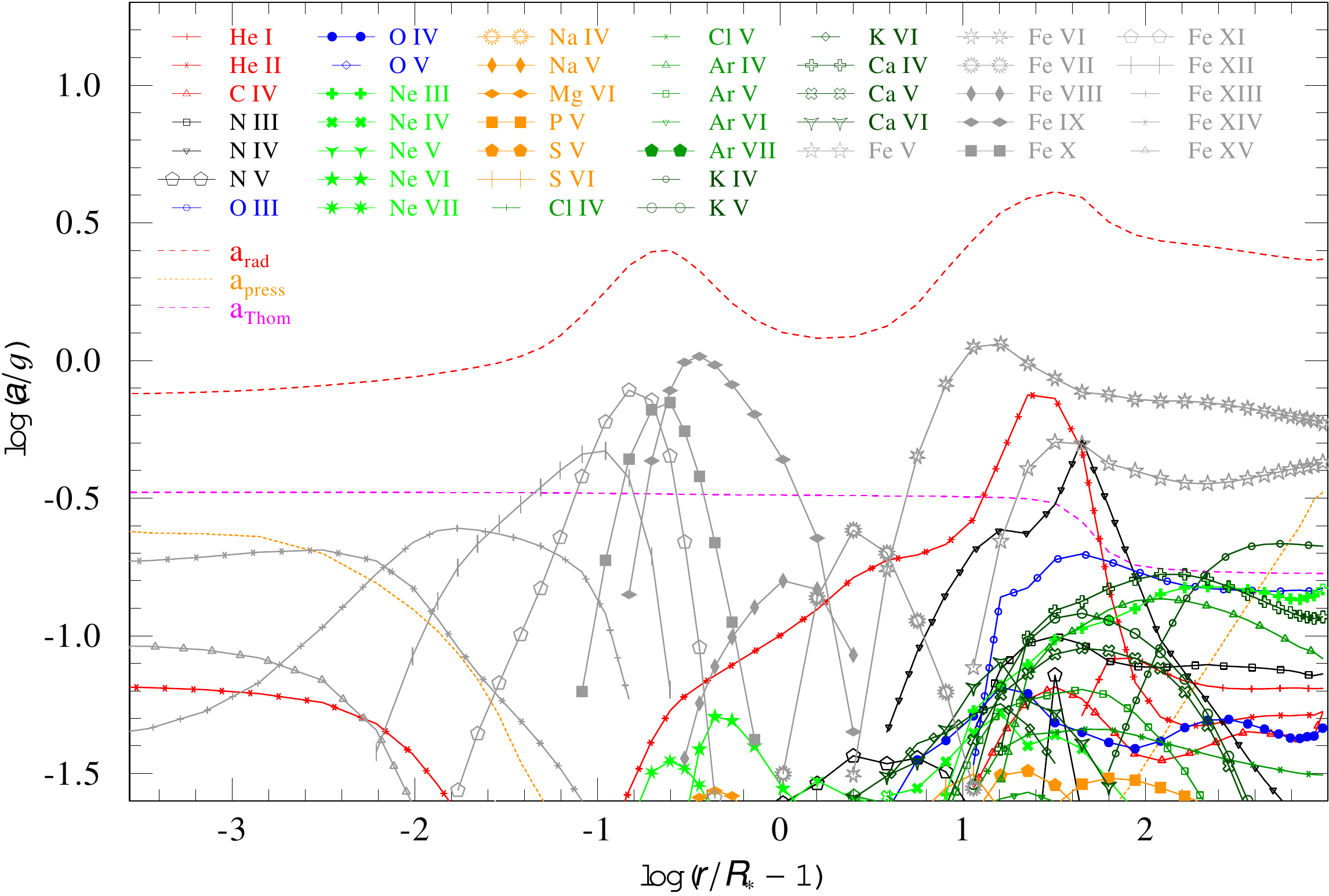}
  \caption{Contributions of the main driving ions to the radiative acceleration $a$ (in function of Newtonian gravity $g$) of a hydro-dynamically consistent N-type WR model with 
$T = 141\,$KK from Sander et al. (2020). 
The main contributions are shown with different ions indicated by a combination of different colors and symbols. 
For comparison the total radiative acceleration ($a_{\rm rad}$), the Thomson acceleration from free electrons, and the contribution from gas (and turbulent) pressure are also shown.}
  \label{f_leadions-wne}
\end{figure}

The crucial point of such an optically thick wind analysis  
is that due to their huge mass-loss rates, the atmospheres become so 
extended that the sonic point of the wind is already reached 
at large flux-mean optical depth, which implies 
that the radiation can be treated in the diffusion approximation. 
The equation for the radiative acceleration can then be approximated to:

\begin{equation}
 \label{eq:arad}
  a_{\rm rad} = \frac{1}{c} \int \kappa_\nu F_\nu {\rm d}\nu
  \simeq \kappa_{\rm Ross}\frac{L_\star}{4\pi r^2 c},
\end{equation}
where $\kappa_{\rm Ross}$ is the Rosseland mean opacity (which can be
taken from the OPAL opacity tables of Iglesias \& Rogers 1996). 
The Eddington limit with respect to the Rosseland mean opacity is being
crossed at the sonic point.
Nugis \& Lamers (2002) showed that the condition that $\kappa_{\rm Ross}$ needs to increase
outward with decreasing density could be fulfilled at the hot edges of 
two Fe opacity peaks, the ``cool'' bump at $\sim$ 70\,kK and the ``hot'' bump 
above 160\,kK (which can be identified as the 2 bumps in Figs.\,\ref{f_checkopaross} and \ref{f_leadions-wne}).  

Instead of $a_{\rm rad}$, it is probably more insightful to express the radiative acceleration $a_{\rm rad}(r)$ through the Eddington ratio
\begin{equation}
  \label{eq:gammarad}
	\Gamma_{\rm rad}(r) = \frac{a_{\rm rad}(r)}{g_{\rm Newton}(r)} = \frac{\kappa_{F}(r)}{4\pi c G}\frac{L_{\ast}}{M_{\ast}}
\label{eq_gamma}
\end{equation}
where the radial dependency of $\Gamma_{\rm rad}$ is due to the flux-weighted mean opacity $\kappa_{F}(r)$. The quantity $\kappa_{F}$ should not be confused with the Rosseland opacity 
that is used in stellar structure calculations or wind initiation studies (Nugis \& Lamers 2002; Ro \& Matzner 2016; Gr\"afener et al. 2017, Grassitelli et al. 2018, Poniatowski et al. 2021). 
In the deeper layers of the atmosphere where the diffusion approximation is valid, $\kappa_{F} \approx \kappa_{\rm Ross}$ as shown in Fig.\,\ref{f_checkopaross}, but the difference between the two quantities becomes significant further out in the wind where they become discrepant by more than an order of magnitude. Therefore, the construction of a consistent wind stratification boils down to an accurate calculation of $\kappa_{F}(r)$ through the entire atmosphere including both the deep optically thick regions and the outer optically thin ones.

Gr\"afener \& Hamann (2005), Sander et al. (2020) and Sander \& Vink (2020) included the Opacity Project 
Fe-peak opacities from the ions Fe\,{\sc ix--xvii} in sophisticated PoWR models for classical WR stars. An example from Sander et al. (2020) for a 
hydrodynamically consistent model is shown  in Fig.\,\ref{f_leadions-wne}, where the radiative acceleration has 
been dissected in individual ions, and compared to other accelerations relevant in the equation of motion. 
It is clear that Fe is most important in the inner wind where the mass-loss rate is set, while the 
opacities of CNO and intermediate-mass elements, such as Ne, S, Ar, Cl dominate 
the outer wind (Vink et al. 1999).

It should be noted that wind clumping in these sophisticated non-LTE models is treated in the optically thin ``micro'' clumping 
approach (see Sect.\,\ref{sec_micro}). For Monte Carlo models both the optically thin and thick approaches have been studied (Muijres et al. 2011). 
For less ad-hoc approaches, multi-D hydrodynamical approaches such as those modelling the line-deshadowing instability (LDI) 
are required (Sundqvist et al. 2018; Driessen et al. 2019; Lagae et al. 2021).
The point of the LDI physics is that a small 
perturbation in velocity leads -- via $a_{\rm rad} \propto \delta v$ -- to a further velocity enhancement, i.e. a runaway effect, usually referred to as the self-excited LDI (Owocki \& Rybicki 1984).

\subsubsection{The Transition mass-loss rate}
\label{sec_trans}

There are many uncertainties in the quantitative 
mass-loss rates of both VMS as well as canonical 20-60\,\msun\ massive stars. 
One reason is related to the role of wind clumping, which will be discussed later, but 
the most pressing 
uncertainty is actually still {\it qualitative}:
do VMS winds become optically thick? 
And if they do, could this lead to an accelerated 
enhancement of $\dot{M}$? And at what location does this transition occur?

To find this transition point and its associated mass-loss rate Vink \& Gräfener (2012) performed 
a full analytical derivation, of which I will here only discuss some key aspects. 
The basic point is to solve the equation of motion (Eq.\,\ref{eq_eom}) in integral form, and 
assuming that hydrostatic equilibrium is a good approximation 
for the subsonic part of the wind, while in the supersonic regime the gas pressure 
gradient becomes small (see Abbott 1982). 
Employing the mass-continuity equation, 
and the wind optical depth $\tau = \int_{r_s}^\infty \kappa_{\rm F}\rho\, {\rm d}r$, one obtains

\begin{equation}
\frac{\dot{M}}{L/c} dv = \kappa_{\rm F} \rho \frac{\Gamma - 1}{\Gamma} dr = \frac{\Gamma-1}{\Gamma} d\tau.
\label{eq_gaga}
\end{equation}
Assuming for O-stars that $\Gamma$ is
significantly larger than 1 in the supersonic region, such that the factor
$\frac{\Gamma-1}{\Gamma}$ becomes close to unity, one can show that

\begin{equation}
\eta = \frac{\dot{M} v_{\infty}}{L/c} = \tau .
\label{eq_eta}
\end{equation}

One can now employ the unique condition $\eta = \tau = 1$ 
exactly at the transition
from optically thin winds to optically-thick WR winds. 
In other words, if one were to have a full data-set listing all   
luminosities for 
O and WNh stars, the corresponding transition mass-loss rate $\dot{M}_{\rm trans}$ is obtained 
by simply considering
the transition luminosity $L_{\rm trans}$ and the terminal velocity $v_{\infty}$:

\begin{equation}
\dot{M}_{\rm trans} = \frac{L_{\rm trans}}{v_{\infty} c}
\label{eq_transm}
\end{equation}
This transition mass-loss rate can be obtained by purely spectroscopic means, 
{\em independent} of any assumptions regarding wind clumping.

Vink \& Gr\"{a}fener (2012) followed a model-independent approach, 
adopting $\beta$-type velocity laws, as well as 
full hydrodynamic models, computing the 
integral $\tau = \int_{r_s}^{\infty} \kappa \rho\,{\rm d}r$  
numerically using the flux-mean opacity $\kappa_{\rm F}(r)$. 
The mean opacity $\kappa_F$ follows from the resulting
radiative acceleration $a_{\rm rad}(r)$.
The sole assumption entering the transition mass-loss analysis is that the winds 
  are radiatively driven. 
The resulting mean opacity $\kappa_F$ captures all physical effects that could affect
  the radiative driving, including clumping and porosity.
The obtained values for the correction factor  
is 0.6 $\pm$ 0.2. The transition between O and WNh
spectral types should in reality occur 
at $\dot{M} = f \frac{L_{\rm trans}}{v_{\infty} c} \simeq 0.6 \dot{M}_{\rm trans}$.
(Vink \& Gr\"afener 2012).

For the Arches cluster the spectroscopic transition occurs at  $\log(L)=6.05$ and
  $\log(\dot{M}_{\eta=1}/\msunyr)=-4.95$ (Martins et al. 2008).
This is related to the transition mass-loss rate for Galactic metallicity (Vink \& Gr\"afener 2012). 
The only uncertainties result from relatively small uncertainties in the 
terminal wind velocity (of about 10\%) and the stellar luminosity $L$ (30\% at worst), 
with potential errors of $\sim$40\% at most, these are several orders of magnitude lower 
than the humongous uncertainties in the accuracy of empirical mass-loss 
rates resulting from the dual effects of clumping and porosity, which could easily 
amount to an order of magnitude or more (e.g. Fullerton et al. 2006).

\begin{figure}
\includegraphics[width=15cm,height=11cm]{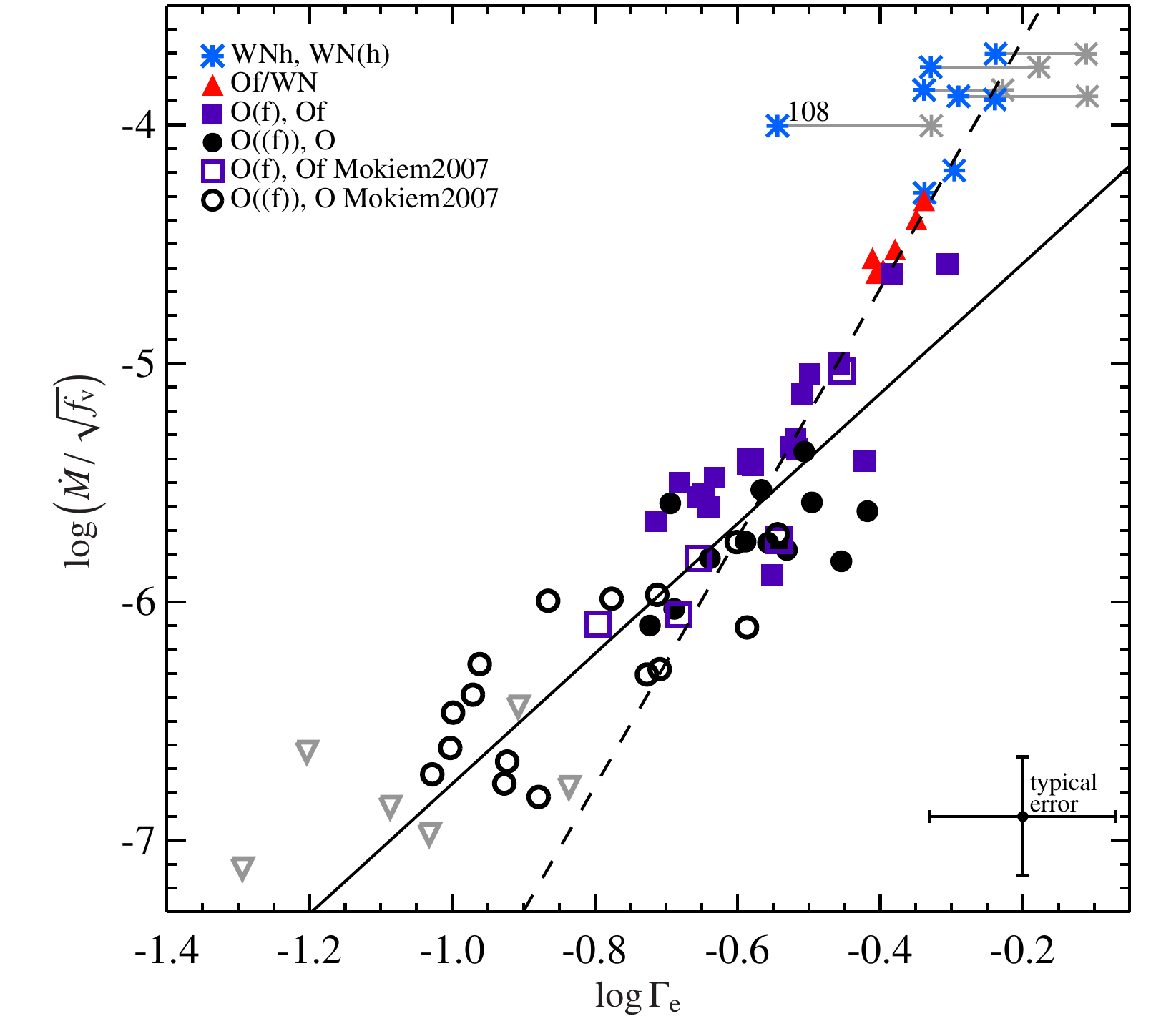}
\caption{Empirical $\dot{M}$-$\Gamma$ relations from Bestenlehner et al. (2014)  -- with similar slopes to those of 
the predicted mass-loss "kink" (Vink et al. 2011). The mass-loss rates are unclumped and could be corrected using the volume filling factor $f$).
$\Gamma$ only accounts for electron scattering. 
Symbols highlight different stellar sub-classes.
The solid line indicates the O-star slope ($\simeq 2$), while the steeper ($\simeq 5$) dashed line covers the very massive Of/WN and WNh stars. 
The Vink et al. (2011) {\it kink} occurs at $\log
  \Gamma_{\rm e} = -0.58$. The grey asterisks indicate the
    position of the stars under the assumption of
    core He-burning. The grey upside down triangles are from
  Mokiem et al. (2007) with only upper limits (they are excluded from the fits).}
\label{f_mdotgamma}
\end{figure}

\subsection{Theoretical mass-loss results}

In CAK-type O-star models, the terminal wind velocity scales with a constant value of about 2-3 times the escape velocity. 
For this reason, terminal wind velocities are often discussed in terms of the stellar effective escape velocity.
Moreover, as the escape velocity drops with larger radii during redwards evolution, the wind velocity is anticipated to drop accordingly. Table 1 shows that to first order this is indeed the case, lower \teff\ types have generally lower terminal wind speeds, but this is only part of the story. Important modifications to this general principle occur when stars approach the Eddington limit, or encounter temperature transitions where the opacity changes drastically, such as at BS-Jumps.

\begin{table}[h]
\tabcolsep7.5pt
\caption{Typical wind parameters for different stellar Types}
\label{tab1}
\begin{center}
\begin{tabular}{@{}l|c|c|c|c|l}
\hline
Type   & $T_{\rm eff}$ & $M$           & $v_{\infty}$ & $\dot{M}$   & SN Type\\
       &  (kK)         & ($M_{\odot}$) & (km/s)       & ($\msunyr$) & (speculative)\\
\hline
O        & 30-45       &  20-60        & 2000-3500    & $10^{-7}-10^{-5}$ &  -        \\
WNh      & 35-50       &  80-300       & 1500-3000    & $10^{-4}$         &  -          \\
BSG      & 15-25       &  15-30        & 500-1500     & $10^{-7}-10^{-5}$ &  IIb/IIP-pec \\
YSG      &  5-10       &  10-25        & 50-200       & $10^{-6}-10^{-4}$ &  IIb  \\
RSG      &  3-5       &   10-25        & 10-30        & $10^{-7}-10^{-4}$ &  IIP/IIL \\
LBV low-L & 10-15     &   15-25        & 100-200          & $10^{-5}$         &  IIb \\
LBV high-L & 10-30     &   40-        & 200-500          & $10^{-4}-10^{-3}$  &  IIn  \\
cWR        & 90-200     &  10-30        & 1500-6000    & $10^{-5} - 10^{-4}$ & Ic \\
Stripped He  & 50-80   &  1-5          & 1000         & $10^{-8}$          & Ib \\
\hline
\end{tabular}
\end{center}
\begin{tabnote}
The numbers in this Table should only be taken as typical, with the terms referring to broad evolutionary groupings. 
They are not used in a strict spectroscopic sense. 
E.g. very late WN-type stars such as WN10 here could fall in the "BSG" category despite having a WR emission-line spectrum.
\end{tabnote}
\end{table}

\subsubsection{O-type stars}

The currently most used recipe in stellar evolution modelling are the Monte Carlo predictions of Vink et al. (2000) with an associated $Z$ dependence from Vink et al. (2001). 
There are several aspects to this recipe that form active areas of research. 
One of them involves the absolute levels of predicted mass-loss rates. 
While these Monte Carlo mass-loss rates are already a factor 2-3 lower than
unclumped empirical mass-loss rates (e.g. de Jager et al. 1988) we will later see that some current analyses (Hawcroft et al. in prep.; Brands et al. in prep.) derive clumping factors larger than 4-10. 
For these reasons, new CMF models have been published in recent years that no longer employ the Sobolev approximation. 
These new hydrodynamical CMF models generally predict mass-loss rates another 
factor 2-3 lower than the Vink et al. (2000) rates. 

One example involves the radiation-driven wind predictions with the {\sc fastwind} code by Bj\"orklund et al. (2021) 
where mass-loss rates are a factor 2-3 lower than Vink et al. (2000), although the mass-loss rate predictions depend significantly on the assumed amount of micro-turbulence.
While the $Z$ dependence is similar to Vink et al. (2001), a \teff\ dependence for O-type stars is not present in this prescription. 
Also, the rapid increase of $\dot{M}$ -- by a factor of about 5 in Vink et al. (1999) -- at the bi-stability location of 
20-25 kK is not predicted by the Bj\"orklund et al. approach. 
The reason for the absence of a mass-loss jump has been relayed (Bj\"orklund; priv. comm.) to be a 
lack of radiative opacity at the critical point.

This is reminiscent of a potentially related\footnote{There is no a priori reason to assume that both issues should have exactly the same resolution. While one might be physical, the other might be purely numerical. The key is to figure out which it is.} issue identified by Muijres et al. (2012) for low-luminosity weak O-type winds in the so-called "weak-wind regime" (see also Lucy 2010). 
One possible solution is that the lack of acceleration at the critical point results in genuinely weak or even non-existent winds in these critical regimes. Another possibility is that Nature somehow manages to find a way to push the gas through the critical point with a "canonical" wind solution establishing itself regardless. 
Vink \& Sander (2021) suggested an empirical test with ULLYSES\footnote{https://ullyses.stsci.edu} 
spectra to investigate the weaker, faster wind solutions in more detail.

\begin{marginnote}
\entry{ULLYSES}{Ultraviolet Legacy Library for Young Stars as Essential Standards. Hubble Space Telescope (HST) Director's programme involving 1000 HST orbits. 500 orbits are dedicated to Young Pre-main Sequence stars, and 500 orbits to massive stars in low-$Z$ galaxies. The massive star sample includes approx 250 massive stars.}
\end{marginnote}

\subsubsection{Very massive stars}

In addition to canonical O stars in the 20-60\msun\ range, we switch our attention to the more massive stars.
Vink et al. (2011) discovered a kink in the slope of the
mass-loss vs. $\Gamma$ relation at the transition from optically thin O-type
to optically thick WNh-type winds, while Bestenlehner et al. (2014) performed a homogeneous spectral 
analysis study of over 60 O-Of/WN-WNh stars in 30 Doradus as part of the VLT Flames Tarantula Survey (VFTS). They 
confirmed the predicted kink empirically, although one could instead modify CAK  
parameters as an alternative (Bestenlehner et al. 2014; Bestenlehner 2020). 

\begin{marginnote}
\entry{30 Doradus}{Largest H{\sc ii} region nearby. Also called the Tarantula Nebula located in the LMC. The region hosts dozens of VMS as well as the most massive young stellar cluster R136, which was once considered a supermassive star of over 1000\,\Msun. While high-resolution imaging has shown the cluster to contain more than just one object, the region still hosts the most massive stars currently known of up to 200-300\,Msun\ (Crowther et al. 2010; Martins 2015).} 
\end{marginnote}

Figure\,\ref{f_mdotgamma} depicts empirical mass-loss predictions for VMS 
as a function of the Eddington parameter $\Gamma$ from Bestenlehner et al. (2014). 
For ordinary O stars with relatively ``low'' $\Gamma$ the $\dot{M}$ $\propto$ 
$\Gamma^{x}$ relationship is shallow, with $x$ $\simeq$2. 
There is a steepening at higher $\Gamma$, where
$x$ becomes $\simeq$5 (Vink et al. 2011). 
Alternative models for VMS have been presented by Gr\"afener \& Hamann (2008) for stars up to 150\,\msun\, and Pauldrach et al. (2012) and Vink (2018a) for stars up to 1000\msun. 

\begin{figure*}
 \includegraphics[width=0.9\textwidth]{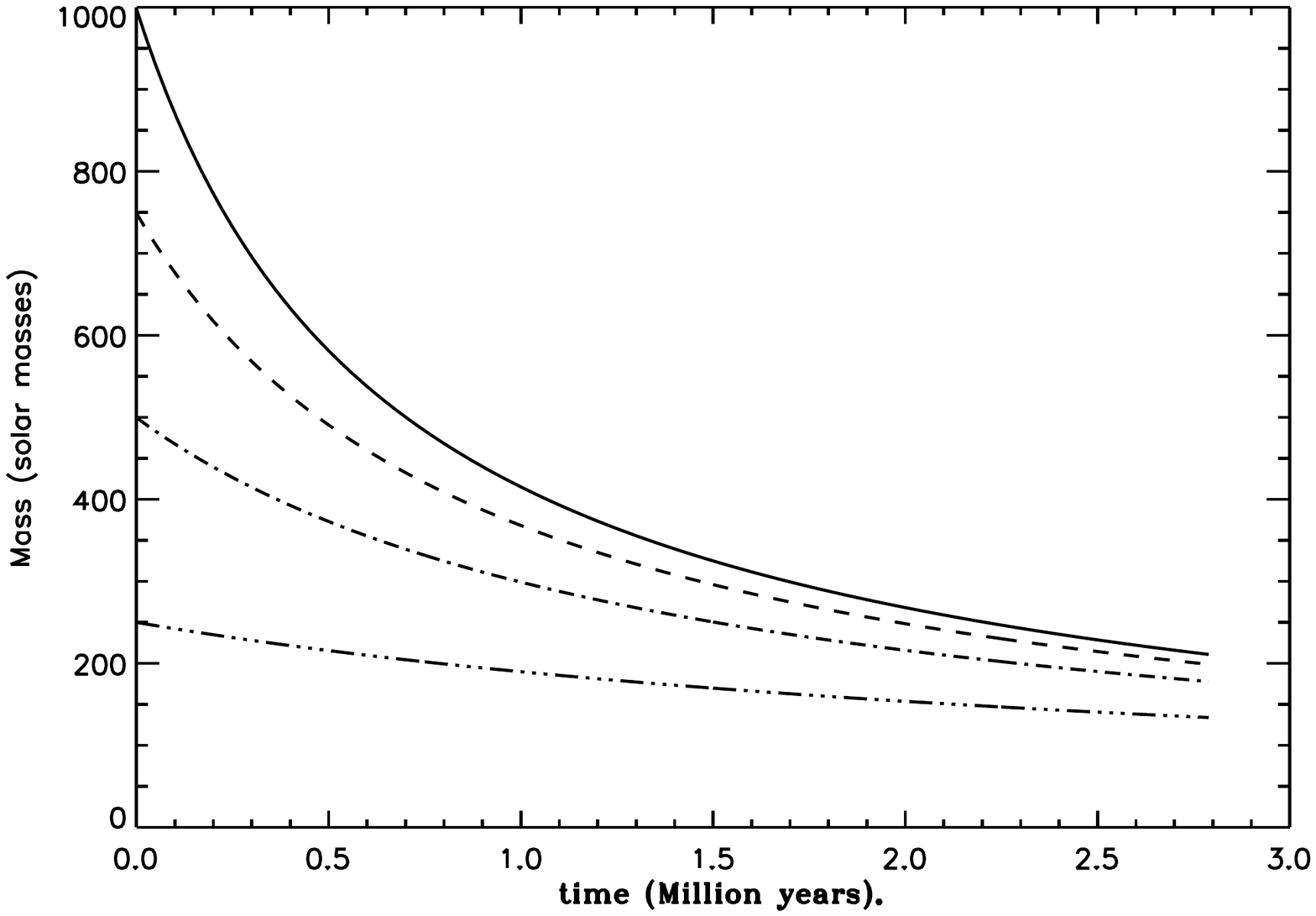}
\caption{Simplified mass evolution for the most massive stars. The effective upper-mass limit  
drops with time, as four different initial masses (250, 500, 750, 1000 \msun) -- represented by the 
four different lines -- `converge' to very similar values after just a few million years.
The plot highlights that it is not possible to determine the value of the upper-mass limit from
present-day stars up to approx 200\,\msun\ unless we have exact knowledge of the mass-loss history. 
From Vink (2018).}
\label{f_evap}
\end{figure*}

Whether and at what stage the winds of VMSs become optically thick (Gr\"afener \& Hamann 2008) or not (Pauldrach et al. 2012) is still debated and an active field of research.
The mass-loss history is crucial 
for understanding the stellar upper mass limit (Vink 2018), as currently observed VMS may have originated from a range of initial stellar masses. 
The reason is the rapid drop in mass already in their first few million years as depicted in Fig.\,\ref{f_evap}. 
Basically, VMS evaporate extremely rapidly if the mass-loss rates are higher than $10^{-4}$ \msunyr.

\subsubsection{Blue Supergiants and the bi-stability jump(s)}

With respect to B supergiants (Bsgs) it should be noted that the evolutionary state of these stars is still under debate. They might either already be burning He in their cores, or they might be objects on an extended main sequence due to a large amount of core overshooting with $\alpha_{\rm ov} = 0.3-0.5$ (Vink et al. 2010; Castro et al. 2014; Higgins \& Vink 2019). 
Their winds are slower than those of their O-star counterparts, dropping by a factor of approx 2 at spectral type B1 with an effective temperature of about 21 kK (Lamers et al. 1995; Crowther et al. 2006). 
This drop is usually referred to as the BS-Jump (Pauldrach \& Puls 1990). It has been attributed to a change of the main wind driver
Fe from {\sc iv} to Fe {\sc iii} (Vink et al. 1999).  In the original MC models the wind efficiency number $\eta$ was found to increase by a factor 2-3, and with an assumed drop in $v_{\infty}$ by a factor of approx 2 this implied an increase in the mass-loss rate of about 5. Contrary to the original global approach, the drop in $v_{\infty}$ has more recently been predicted directly with locally consistent models (e.g. Vink \& Sander 2021).
Despite the important consequences for massive star evolution models, for instance in terms of angular momentum loss (Vink et al. 2010; Keszelthy et al. 2017), the predicted $\dot{M}$ increase has sofar hardly been tested. This has become all the more pivotal now that new CMF modelling with {\sc fastwind} by Bj\"orklund et al. (2021) exhibit the complete lack of an increased $\dot{M}$ at the bi-stability jump. 

So far, empirical wind diagnostics tests have only been performed in \ha\ (Crowther et al. 2006; Markova \& Puls 2008) and the radio regime (Benaglia et al. 2007; Rubio-Diez et al. 2021). 
While the first evidence for a local maximum in mass-loss rate at spectral type B1 was identified, empirical radio and \ha\ rates below the bi-stability jump \teff\ are up to an order of magnitude lower than predicted. 
It still needs to be established exactly what happens at spectral type B1 when Fe {\sc iv} recombines to Fe {\sc iii}.
One possibility is that when the mass-loss rate increases as predicted this leads to optically thicker winds that have different diagnostics (see Sect.\,\ref{sec_macro}) as suggested by Petrov et al. (2014). Another possibility is that the MC calculations are inaccurate and that the recently predicted absence of a BS-Jump by Bj\"orklund et al (2021) turns out to be more accurate. One should however note that very low mass-loss rate predictions would normally go hand-in-hand with very fast winds, which are not observed. In fact the empirical drop in $v_{\infty}$ is well established, whether it is abrupt (Lamers et al. 1995) or more gradual (Crowther et al. 2006) is a different matter. 

To make progress on the issue of the reality of the BS-jump it would be helpful if we were able to devise a testing strategy that is model-independent, as so far the issue is that when theory changes abruptly at the BS-jump, so does the diagnostics. A promising alternative empirical method might involve the bow-shock method 
(Kobulnicky et al. 2019) discussed in Sect.\,\ref{sec_bow}.

\subsubsection{Cool Supergiants}

When considering cooler objects it should be noted that another model prediction is that of a second BS-Jump at \teff\ lower than approx 10 kK
(Vink et al. 1999; Petrov et al. 2016). This is caused by a recombination of Fe {\sc iii} to Fe {\sc ii} and its existence and size are even more uncertain than the first BS-jump, noting that also in this \teff\ range -- corresponding to spectral type A0 -- Lamers et al. (1995) found a rapid drop in terminal wind speeds by a factor 2 empirically in IUE spectra.

The cool supergiants (and hypergiants) could be divided between red (3-5 kK) and yellow (5-10 kK) supergiants. The driving mechanism of RSGs is usually assumed to be radiation pressure on dust\footnote{But this is far from settled, see e.g. Kee et al. (2021).} 
while hot-star winds above 10 kK are driven by radiation pressure on gas.
The driving mechanism for yellow supergiants (YSGs) is still unknown but these objects find themselves at a critical HRD location, right at the boundary between the convection and/or dust-driven 
RSGs, and the line-driven BSGs. From an evolutionary perspective YSGs also play an important role, possibly as post-RSGs, which might imply their death could be imminent. While convection, pulsation, and radiation pressure might all play a role in the driving of YSG winds, the theory is in dire need for further development (Lobel et al. 2003; Andrews et al. 2019; Koumpia et al. 2020) as their mass-loss rates are critical for understanding the HD limit (Gilkis et al, 2021; Sabhahit et al. 2021).

\subsubsection{WR and stripped He stars}

The first sophisticated theoretical model for a WR wind did not appear 
until 2005 (Gr\"afener \& Hamann), which is why the stellar evolution community still had to rely on empirical mass-loss rates such as those of Nugis \& lamers (2000). More recently, the first grids of hydro-dynamically consistent PoWR models have been published (Sander \& Vink 2020). 

\begin{figure}
 \includegraphics[width=0.9\columnwidth]{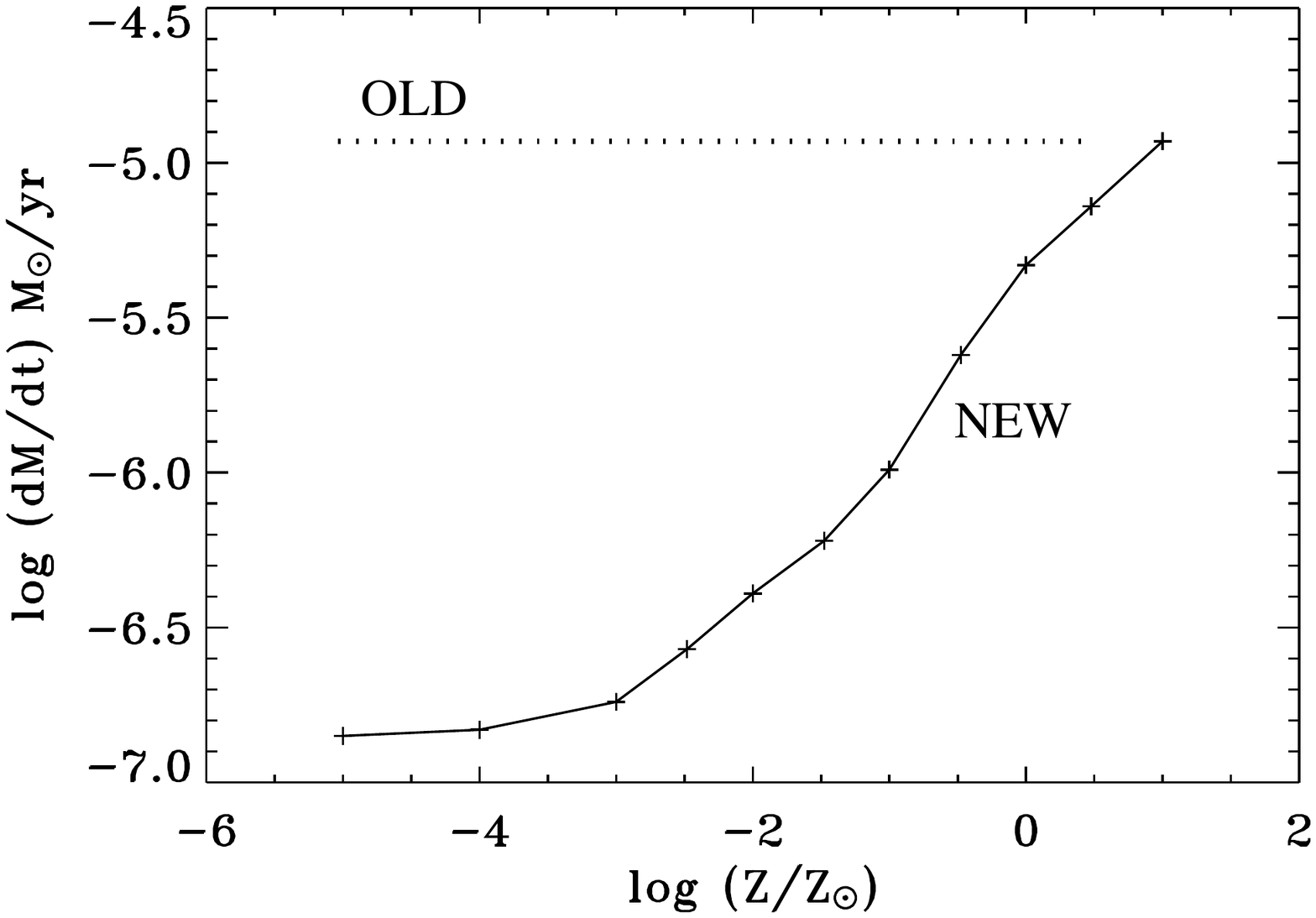}
 \caption{$Z$-dependent mass-loss predictions for WR stars.
Despite the overwhelming presence of C over the full $Z$ range, the Fe-dependent WR mass-loss rate does {\it not} show
$Z$-independent behaviour, as was previously assumed (dotted line).
The newer computations show that WR mass loss depends on Fe --- a {\it key}
result for predicting a high occurrence of long-duration GRBs at low $Z$.
At very low $Z$, the slope becomes smaller, flattening off completely at $Z/\Zsun$ $=$ $10^{-3}$. 
The MC computations are from Vink \& de Koter (2005).}
  \label{f_mdotz2}
\end{figure}

During the WR phase, the atmospheres become chemically enriched with elements such as carbon (C), which may potentially modify WR mass loss. 
In a first attempt to investigate the effects of self-enrichment on the
total wind strength, Vink \& de Koter (2005) performed a pilot study of late-type 
WR mass loss versus $Z$. 
The key point here is that WR stars,
especially C-types (WC stars), show the products of core He burning in their
outer layers, and it could have been the C abundance that is 
most relevant for driving WC winds rather than Fe.
Figure~\ref{f_mdotz2} however shows
that -- despite the fact that the C ions overwhelm the amount of Fe -- WR stars 
show a strong $\dot{M}$-$Z$
dependence, because Fe has a more complex atomic structure involving 
millions of spectral lines.

The implications of Fig.~\ref{f_mdotz2} are two-fold. 
First, WR mass-loss rates decrease steeply with $Z$. 
This may be of key relevance for black hole formation and
the progenitor evolution of long duration gamma-ray bursts (GRBs). 
The collapsar model of Woosley (1993) requires a rapidly rotating stellar core prior to 
collapse, but at solar metallicity stellar winds are expected to remove the bulk of 
the angular momentum. The WR $\dot{M}$-$Z$ dependence from
Fig.~\ref{f_mdotz2} provides a route to maintain rapid
rotation, as the winds are weaker at lower $Z$, even during the final phases towards 
collapse.

The second point is that mass
loss does not decrease when $Z$ falls 
below $\sim$$10^{-3} \Zsun$, resulting from 
the dominance of radiative driving by carbon lines. 
This suggests that even for Pop III stars with only trace amounts of metals, 
once the heavy objects enrich their outer atmospheres, radiation-driven 
winds might still exist.  
Whether the mass-loss rates are sufficiently high to alter the
evolutionary tracks of these First Stars remains to be seen, but it 
is important to keep in mind that the mass-loss physics does not 
only depend on $Z$, but that other 
factors, such as the proximity to the $\Gamma$ limit, could also be relevant.

\begin{figure}
  \includegraphics[width=\textwidth]{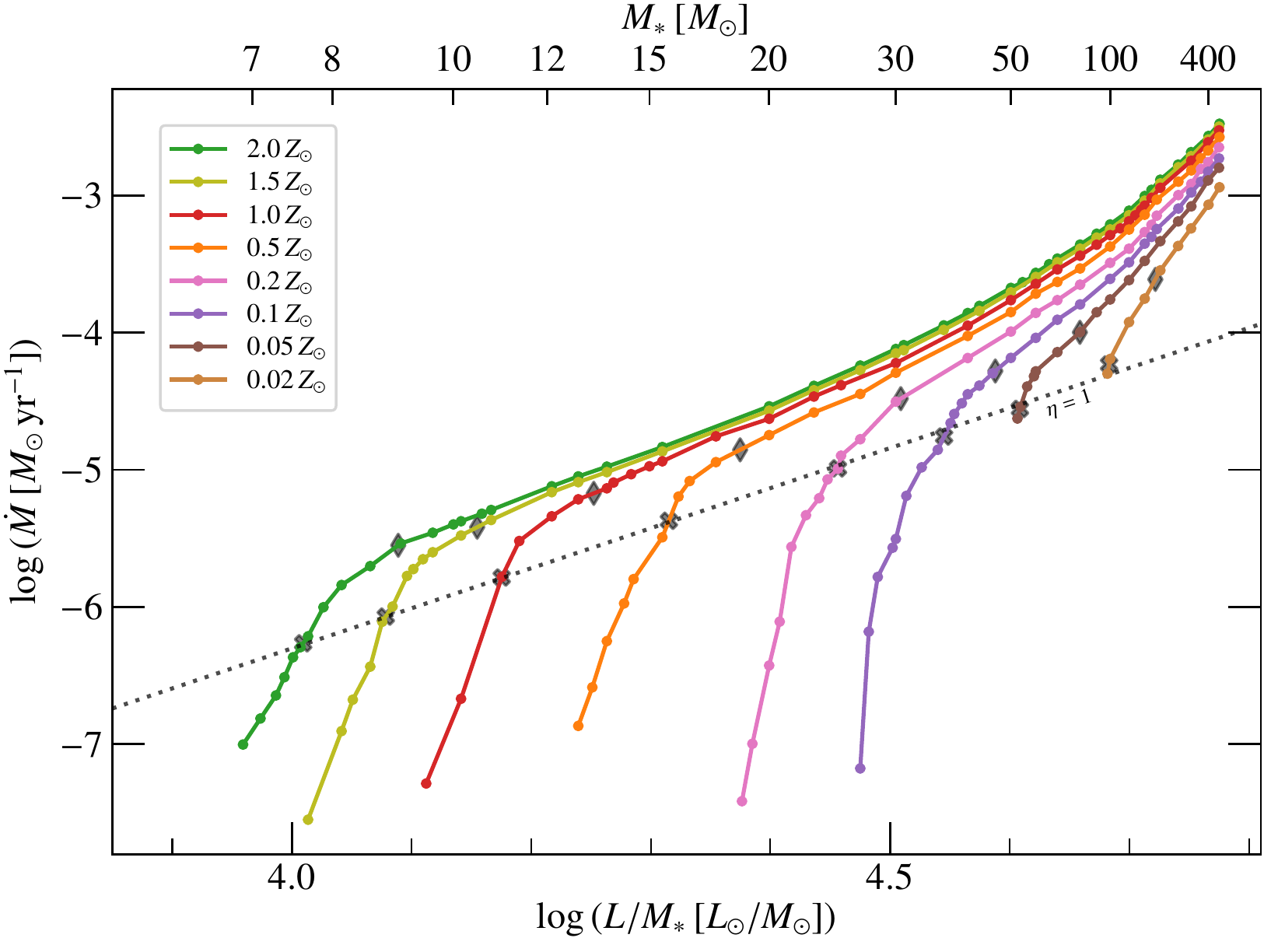}
  \caption{Mass-loss rate $\dot{M}$ as a function of $L/M$ for sets of hydro-dynamically consistent atmospheres for a wide range of $Z$ from Sander \& Vink (2020). 
The onset of multiple scattering $\eta = 1$ in each sequence is marked with a grey cross and a dotted line. Grey diamonds denote the locations of $\eta = 2.5$.}
  \label{f_mdotWR}
\end{figure}

From more recent computational results it has also become clear that the \mdot\ dependence on WR parameters such as $M$ is not a power-law either. 
Figure\,\ref{f_mdotWR} showcases the combined $M$ and $Z$ dependence of N-type WR stars from hydrodynamical self-consistent models (Sander \& Vink 2020). Interestingly, the figure also shows a $Z$-dependent break-down where the winds become optically thin. 
This region appears to correspond to that of the "Stripped He stars" (e.g. G\"otberg et al. 2020) as these He stars -- of just a few solar masses -- have significantly weaker winds than the optically thicker WR winds. 
The transition luminosity for different $Z$ galaxies between these optically thinner and thicker stars was discussed by Shenar et al. (2020). Note that the predictions in Fig.\,\ref{f_mdotWR} have sofar only been computed for a single temperature and the steep drop should therefore not be taken too literally. For now it might be recommendable to use the Monte Carlo predictions of Vink (2017) as a lower bound (see Appendix).

Future hydrodynamic mass-loss calculations and observations are warranted for this He-star regime, as the SN-type ultimately produced (ie. a Type Ib, Type IIb, etc.) 
depends very critically on the included mass-loss rate, given that there is only a very small amount of H on top, and extrapolations from optically thick empirical WR recipes may not be justified (Vink 2017; Gilkis et al. 2019).

\begin{summary}[SUMMARY POINTS]
\begin{enumerate}
\item The traditional CAK line-driven wind theory is being updated with Monte Carlo and non-Sobolev co-moving frame models that provide $\dot{M}$ as a function of $M$ and $L$, as well as $T_{\rm eff}$ and appropriate scalings on the Fe abundance. 
\item These new models provide the velocity stratification $v(r)$ showing that traditional $\beta$ law approaches need to be re-evaluated.
\item Different theoretical models, using different assumptions, do not agree on the absolute mass-loss rates (within approx factors 2-3), but for the most massive stars Monte Carlo rates are of the right order of magnitude, as determined from calibrations using the transition mass-loss rate.
\end{enumerate}
\end{summary}

\section{Mass-loss Diagnostics}
\label{sec_obs}

RDWT for smooth winds generally provides predictions for 
two global wind parameters: \mdot, and $v_{\infty}$.
Most diagnostics are based on the wind density, and the mass-loss rate then simply follows from the continuity equation.
Up until recently, most diagnostics have been based on wind models that employ a pre-scribed $\beta$-type velocity field, but with new 
hydrodynamical consistent CMF approaches this is no longer necessary. 
 Regarding the global wind parameters $v_{\infty}$ is directly observable from P\,Cygni absorption profiles (with a 10\% accuracy) and for emission line WR stars the width of the line may also provide relatively direct wind velocity information. Note that this does not include the very broad wings which are caused by electron scattering and which provide interesting constraints on the amount of wind clumping in WR stars (Hillier 1991). 

The mass-loss diagnostics is generally far more model-dependent, although the transition mass-loss rate gives a model-independent value in the very high-mass range around 80-100\,$M_{\odot}$. Ideally we would be able to derive accurate model-independent mass-loss rates in the canonical 20-60\,$M_{\odot}$ range as well. One recently proposed method is that of the bow-shock method (see Sect.\,\ref{sec_bow}).

An important issue to keep in mind when comparing wind theories to observations is that whenever the theory undergoes discontinuities, such as at BS-Jumps, then the diagnostics might also get affected. For instance, in the case of the BS-Jump temperature it is rather unlikely that the recombined winds on the cool side of the jump can be directly compared to those of the ionised diagnostics on the hot side of the jump (Groh \& Vink 2011). In the case of the \ha\ line Petrov et al. (2014) found that for a reasonable range of parameters the stars on the cool side of the BS-Jump \ha\ might become optically thick, and if the winds are clumped we will see later that while optically thin clumping generally leads to an {\it over}estimate of unclumped empirical mass-loss rates, for optically thick cases the reverse might be true. 
These kind of considerations need to be kept in mind when comparing theory to observations.

\subsection{Multi-wavelength Diagnostics}

Traditionally, the mass-loss rate can be obtained from either stellar continua at long wavelengths (mm and radio; see Sect.\,\ref{sec_radio}) 
or spectral line transitions in the optical or UV (and alternatively also in the X-ray and near-IR domains).
Sticking to the optical and UV wave-bands, there is a profound difference between UV resonance lines 
and (optical) recombination lines such as \ha\ (see the left-hand side versus the right-hand side of Fig\,\ref{f_oski}). 
For resonance lines, the line opacity $\propto \rho$, while recombination involves 
a 2-body process where the line emissivities $\propto \rho^2$. 
This becomes particularly relevant when wind clumping is taken into account in the analysis.
Most empirical analyses rely on the use of non-LTE
model atmospheres. In state-of-art modelling the stellar and wind parameters, such as 
$\dot{M}$, are determined by fitting the resonance and recombination lines simultaneously (e.g. Hillier 2020).

A detailed discussion of the various methods to
derive wind parameters is given in Puls et al. (2008). 
The most common line profiles in a stellar wind are 
UV P\,Cygni
profiles with a blue absorption trough and a red emission peak, and  
optical emission lines such as \ha. 
In P~Cygni scattering lines the upper level is populated by the balancing act 
between absorption from -- and spontaneous decay to -- the lower level. 
An emission line is formed if the upper level is populated by recombinations 
from above (see however Puls et al. 1998; Petrov et al. 2014 for the formation 
of P\,Cygni optical H$\alpha$ lines in the cooler BA supergiants).

\subsubsection{Ultraviolet P\,Cygni Resonance lines}
\label{sec_uv}

P\,Cygni lines have been used to determine the terminal wind velocities $v_{\infty}$ for decades 
(see Prinja et al. 1990 and Lamers et al. 1995 for results with IUE).
UV P-Cygni lines from hot stars (e.g. C\,{\sc iv} and P\,{\sc v}) 
are usually analysed by means of the Sobolev optical depth.
In principle, $\dot{M}$ can be derived from resonance line 
P\,Cygni profiles if the ionisation fraction is known. 
However, for higher $Z$ stars, most P\,Cygni lines are saturated and mass-loss rate derivations become
unreliable such that only lower limits on $\dot{M}$ can be determined. 
UV resonance lines instead have been considered relatively "clean" from wind clumping effects, but 
this would no longer hold if porosity effects become relevant (see Sect.\,\ref{sec_macro}).
The Far UV spectroscopic explorer (FUSE) was instrumental in identifying the Phosphorus {\sc v}
Problem (see Sect.\,\ref{sec_pv}), while the new Director's discretionary project ULLYSES is expected to enable
tremendous progress on wind velocities of hundreds of stars in low $Z$ environments with the Hubble Space Telescope.

\subsubsection{The H$\alpha$ recombination emission line}
\label{sec_ha}

The most oft-used diagnostics to derive \mdot\ for O-star winds 
involves H$\alpha$, for which there is hardly any uncertainty due to 
H ionisation (Klein \& Castor 1978; Drew 1989; Lamers \& Leitherer 1993; Puls et al. 1996). 
As the H$\alpha$ opacity scales with $\rho^2$, 
any notable inhomogeneity will result in an 
\mdot\ overestimate if clumping were neglected in the analysis. 
An advantage of \ha\ is the fact that it remains optically thin in the main part of
the emitting wind, such that porosity effects can be
neglected (which is not the case for UV resonance lines). 
A drawback of the \ha\ line as a mass-loss diagnostic is that it is only sensitive for mass-loss rates above $10^{-7}$ \msunyr\ (Mokiem et al. 2007). For weaker lines the near infrared regime might provide alternative $\rho^{2}$ diagnostics (Najarro et al. 2011).

A dual analysis of UV plus optical lines is hence critical for an accurate multi-wavelength view of stellar winds, and for this reason the massive
star community suggested a complementary optical (and near-IR) counterpart to ULLYSES, called "X-Shooting ULLYSES (XSHOOTU\footnote{https://massivestars.org/xshootu/}), which in conjunction with the UV data is needed to disentangle the effects of clumping and porosity.

\begin{marginnote}
{XSHOOTU}{X-shooting ULLYSES. An ESO-VLT Large Programme with the X-Shooter spectrograph, involving 250 low-$Z$ massive stars, and the optical counterpart to the UV Legacy Library. }
\end{marginnote}

\subsubsection{Radio and (sub)millimetre continuum emission} 
\label{sec_radio}

A complementary approach to measure mass-loss rates is to utilize 
long wavelength thermal radio and (sub)millimetre continua. 
In fact this approach may lead to the most accurate results, as they are considered to be model-independent. 
The basic concept is to measure the excess wind flux over that from the stellar 
photosphere.
This excess flux is emitted by free-free and bound-free processes\footnote{In case of magnetic fields or binary interactions, non-thermal radio emission needs to be taken into account. See Review by De Becker (2007).}. 
The reason the excess flux
becomes more dominant at longer wavelengths is 
thanks to the $\lambda^2$ dependence of the opacities. 
As  
the continuum becomes 
optically thick in the wind in free-free opacity, 
the emitting wind volume increases as a function 
of $\lambda$, leading to the formation of a radio photosphere
where the the radio emission dominates the stellar photospheric emission. 
For a typical
O supergiant this occurs at about 100 stellar radii.
At such large distances the outflow has already reached its terminal velocity,
and an analytic solution of the radiative transfer is achieved, involving the 
spectral index of thermal wind emission close to 0.6 (Wright \& Barlow 1975; Panagia \& Felli 1975).
The method can be used for both sub-mm observations (Fenech et al. 2018) and updated radio facilities 
such as the e-VLA and e-Merlin (Mofford et al. 2020) and should be utilised more frequently.

\subsubsection{X-ray Method}
\label{sec_xray}

The launch of the Chandra and XMM-Newton X-ray telescopes in the last two decades enabled mass-loss rate constraints from 
high-resolution X-ray spectroscopy. 
For a canonical O-star wind one would expect to observe asymmetric line profiles 
(Macfarlane et al., 1991), as the receding wind region attenuates more X-ray line emission than the region in front of the star. This would be expected to result in more line emission on the blue-shifted side than on the red-shifted side, resulting in an asymmetric line profile of X-ray lines.  
However, it turned out that the X-ray profiles were far less asymmetric than expected, suggesting a lower than canonical mass-loss rate (Owocki \& Cohen 2006). However a potential alternative to explain this absence of asymmetry is to entertain the idea that wind clumping reduces the X-ray wind opacity, allowing radiation to escape in a porous medium, and still allow for higher canonical mass-loss rates (Feldmeier et al. 2003). 
Future multi-wavelength mass-loss studies should determine whether the lower opacity is due to an intrinsic mass-loss rate reduction or a porous medium.

A complementary way to study wind clumping in X-rays is in high-mass X-ray binaries (HMXBs) where one can effectively utilise the accreting compact object to probe the clumpy donor wind without the concern of density squared diagnostics (see the recent review in Martinez-Nunez et al. 2017).

\subsubsection{Bow-shock Method}
\label{sec_bow}

A newer method to measure mass-loss rates which is independent of wind clumping is that of the bow-shock method as it principally only depends on the ram pressure between the stellar wind and the ambient interstellar medium (Gvaramadze et al. 2012; Kobulnicky et al. 2019; Henney \& Arthur 2019). 
Following Kobulnicky et al.:

\begin{equation}
\dot{M} = \frac{4 \pi R_{0}^2 v_{\rm a}^2 \rho_{\rm a}}{v_{\rm w}}
\end{equation}
where $R_{0}$ is the angular size of the "standoff" distance where the momentum fluxes of the wind and ISM are equal, $v_{\rm a}$ and $\rho_{\rm a}$ are the velocity and density of the ambient ISM in the stellar rest-frame, and $v_{\rm w}$ is the velocity of the wind. 
$R_{0}$ can be measured from infrared images, while $v_{\rm w}$ can be taken to be the terminal wind speed. While there could be systematic issues with this type of analysis, e.g. due to the difference between the infrared radii and the actual termination shock, or assumptions of adiabatic vs. isothermal shocks, the strength of this new bow-shock method is that it does not depend on wind clumping.

In other words, the method has the potential to provide key insights in wind regimes where there are challenges with traditional diagnostics, such as in the weak-wind regime and on the cool side of the BS-Jump. 

\subsection{Wind clumping}
\label{sec_clumping}

Due to the variability of spectral lines, as well as the presence of 
linear polarisation, we have known for decades that stellar winds are not stationary but time-dependent, and that this leads to
inhomogeneous, clumpy media (see Puls et al. 2008; Hamann et al. 2008).
 
H$\alpha$ and long-wavelength continuum 
diagnostics depend on the density squared, and are thus
sensitive to clumping, whereas UV P Cygni lines such as 
P{\sc v} are insensitive to clumping, as they depend linearly 
on density. 
In the optically thin "micro-clumping" limit, 
the wind is divided into a portion of the wind that 
contains all the material with a volume filling factor $f$, 
whilst the remainder of the
wind is assumed to be void. 
In reality however, clumped winds are porous with a range of clump
sizes, masses, and optical depths, and a macro-clumping approach is needed (Sect.\,\ref{sec_macro}). 

\subsubsection{Optically thin clumping (``micro-clumping'')}
\label{sec_micro}

The general concept of optically-thin micro clumping is  
based on the assumption that the wind is made up of large 
numbers of small clumps. 
Motivated by the results from
hydrodynamic simulations including the line-deshadowing 
instability (LDI; Owocki 2015), the
inter-clump medium is usually assumed to be void. 
The average density is $\rho = f \rho_{C}$,
where $\rho_{C}$ is the density inside the over-dense clumps, and 
the clumping factor $D=f^{-1}$ is a measure for this over-density. 
As the inter-clump space may be assumed to be void, 
matter is only present inside the clumps, with density $\rho_{C}$, and with its 
opacity given by $\kappa=\kappa_C(D \rho)$.
This micro-formalism is only correct as long as the clumps are optically thin, and 
optical depths may be expressed by a mean opacity $\bar \kappa = \kappa_C(D \rho)f = \frac{1}{D} \kappa_C(D \rho)$.  
Hence, for processes that are linearly dependent on density, the mean 
opacity of a clumped medium is exactly the same as for a smooth wind, 
whilst mean opacities are enhanced by the clumping factor D 
for processes that scale with the density squared.

It should be noted that processes 
described by the optically thin micro-clumping approach do not
depend on clump size nor geometry, but only the clumping factor. 
The enhanced opacity
for $\rho^2$ dependent processes implies that $\dot{M}$ 
derived by such diagnostics are a factor of $\sqrt{D}$ lower than
older mass-loss rates derived with the assumption of smooth winds (Mokiem et al. 2007; Ramirez-Agudelo et al. 2017). 

Note that also for the case of thermal radio and (sub)-mm
continuum emission the scaling is exactly the same as for emission lines such as
H$\alpha$. Abbott et al. (1981) showed that the radio flux may be
a factor $f^{-2/3}$ larger than that from a smooth wind with the same
$\dot{M}$, and hence 
radio mass-loss rates derived from clumped winds must also be lower 
than those derived from smooth winds.


\subsubsection{The Phosphorus {\sc v} problem} 
\label{sec_pv}

Due to the very low cosmic abundance of phosphorus (P), the
P {\sc v} doublet remains unsaturated, even when P$^{4+}$ is
dominant. 
This allows for a direct estimate of the product of the mass-loss rate and the ion fraction. 
Unfortunately ion fractions for given
resonance lines can be uncertain due to shocks and associated  
X-ray ionisation processes (Krticka et al. 2009; Carneiro et al. 2016).  
Empirical determination of ionisation fractions is normally 
not feasible, as resonance lines from consecutive ionisation stages 
are not available generally.
Nevertheless, for \PV, insight was obtained from Far-UV data with FUSE. 
For certain O-type ranges the \PV\ line should provide an accurate 
estimate of solely the mass-loss rate, as the pure linear character 
with $\rho$ makes it clumping independent.
Fullerton et al. (2006) selected a large sample of O-stars, which also had 
$\rho^2$ (from H$\alpha$/radio) estimates available,
and compared both $\rho$-linear UV and $\rho$-quadratic dependent 
methods. They found enormous discrepancies, implying 
extreme clumping factors up to D $\sim$ 400 {\it if} the winds could 
be accurately treated in the 
optically thin micro-clumping approach (see also Bouret et al. 2003).

\begin{figure}
\begin{center}
 \includegraphics[width=10cm]{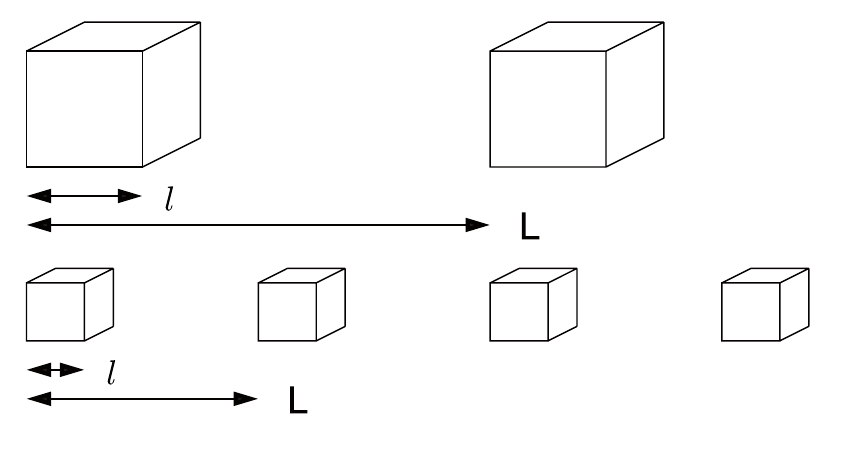}
\end{center}
\caption{Schematic explanation of porosity, involving a notable 
difference between the volume filling fraction f (and its reciprocal clumping factor $D = 1/f$), 
which is the same for the top and bottom case, and the separation of the clumps $L$, which is 
larger in the top case than the bottom case (from Muijres et al. 2011).}
\label{por}
\end{figure}

\subsubsection{Optically thick clumping (``macro''-clumping)} 
\label{sec_macro}

With studies yielding clumping factors $D$ ranging up to 400, one 
may wonder whether a pure micro-clumping analysis is physically sound. 
Most of the atmospheric codes only consider density
variations, but multi-D radiation hydrodynamic
simulations also reveal strong velocity changes inside the
clumps (Owocki 2015). 
Most worrisome is probably the assumption 
that all clumps are assumed to be optically thin. 

Within the optically thin approach, a clump 
has a size smaller than the photon mean free path. 
However, in an optically thick clump, photons may interact with the gas 
several times before they escape through the inter-clump gas. 
Whether a
clump is optically thin or thick depends on the abundance, ionisation
fraction, and cross-section of the transition. 

For optically thick clumps, photons care about the distribution,
the size and the geometry of the clumps (see Fig.\,\ref{por}). 
The conventional description of macro-clumping is based on a  
clump size, $l$, and an average spacing of a statistical
distribution of clumps, $L$, which are related to $f$.
The optical depth across
a clump of size $l$ and opacity $\kappa_C$ becomes:

\begin{equation}
\tau_C = \kappa_C l = \bar \kappa D l = \bar
\kappa \frac{L^3}{l^2} = \bar \kappa h,
\label{eq:taumacroclu}
\end{equation}
with mean opacity $\bar \kappa$ and porosity length
$h=L^3/l^2$. The
porosity length $h$ involves the key parameter to define 
a clumped medium, as $h$ corresponds to the photon mean free path in a
medium consisting of optically thick clumps.
Following Feldmeier et al. (2003) and Owocki \& Cohen (2006)
the effective clump cross section becomes $\sigma_{C} = l^2 \, (1 - e^{-\tau_C})$,
and the effective opacity now becomes:

\begin{equation}
\kappa_{\rm eff} = n_C \sigma_C = \frac{l^2\,(1 - e^{-\tau_C})}{L^3} = 
\bar \kappa \frac{(1 - e^{-\tau_C})}{\tau_C},
\end{equation}
where $n_C$ is the clump number density. 
This equation should be appropriate for clumps of any optical thickness.

\begin{figure}
\begin{center}
\hspace{-0.2cm}
\includegraphics[width=11.5cm]{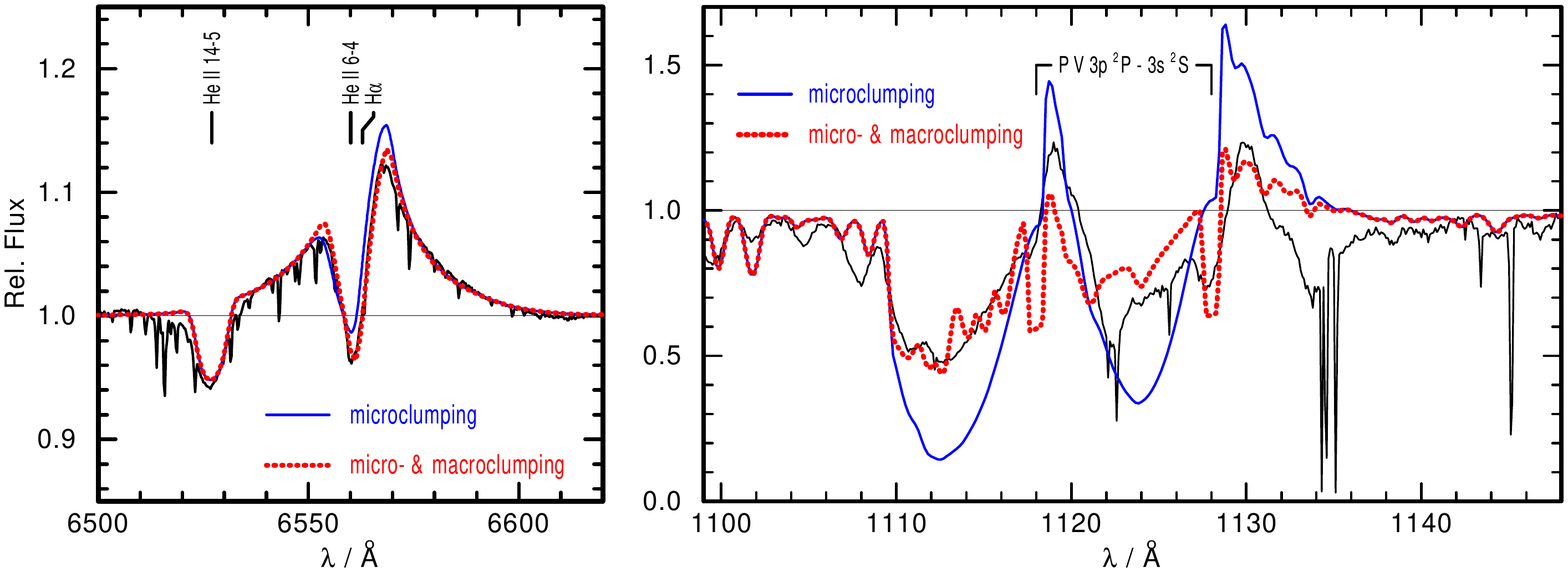}
\vspace{-0.2cm} 
\caption{Porosity as the likely solution for the PV problem. The \ha\ line on the left-hand side is hardly affected by macro-clumping, while the P\,{\sc v} UV line on the right is strongly affected. Adapted from Oskinova et al. (2007).} 
\label{f_oski}
\end{center}
\end{figure}

Oskinova et al. (2007) employed the effective opacity concept 
in the formal integral for the  
line profile modelling of the O supergiant $\zeta$~Pup.
Figure~\ref{f_oski} shows that the most pronounced 
effect involves strong resonance lines, such as \PV\ which can be
reproduced by this macro-clumping approach -- without the need for extremely 
low $\dot{M}$ -- resulting from an effective opacity reduction 
when clumps become optically thick. Given that \ha\ remains 
optically thin for O stars it is not affected by porosity\footnote{This might be different for B supergiants below 
the bi-stability jump (see Petrov et al. 2014).}, and 
it can be reproduced simultaneously with \PV. This enabled
a solution to the \PV\ problem (see also Surlan et al. 2013; Sundqvist \& Puls 2018). 

Note that line processes may also 
be affected by velocity-field changes.
Owocki (2015) showed LDI simulations 
where the line strength was described through a velocity-clumping factor. 
These simulations resulted in a reduced wind absorption due to 
porosity in velocity space, which has been termed ``vorosity''.  
The issue with explaining a reduced \PV\ line-strength
through vorosity is that one needs to have a relatively 
large number of substantial velocity gaps, which does not 
easily arise from the LDI simulations. 
In any case, there is clearly still a need to study
scenarios including both porosity and vorosity, as well as 
how they interrelate (Sundqvist \& Puls 2018).

\subsubsection{The origin of wind clumping}
\label{sec_origin}

In the traditional view of line-driven winds via the LDI, clumping would be expected to develop
in the wind when the wind velocities are large enough to produce shocked
structures. For typical O star winds, this is thought to occur at 
about half the terminal wind velocity at about 1.5 stellar radii.

Various observational indications, including the existence of linear 
polarisation (e.g. Davies et al. 2005) as well as radial dependent 
spectral diagnostics (Puls et al. 2006) however show that clumping should 
already exist at very low wind velocities, 
and more likely arise in the stellar photosphere.
Cantiello et al. (2009) suggested that waves produced by the subsurface 
convection zone could 
lead to velocity fluctuations, and possibly density fluctuations, and 
thus be the root cause for the observed wind clumping at the
stellar surface. 
Assuming the horizontal extent of the clumps to be comparable to the vertical extent in terms of the 
sub-photospheric pressure scale height $H_{\rm p}$, one may
estimate the number of convective cells by dividing the stellar surface area
by the surface area of a convective cell finding that it scales as
($R/H_{\rm P})^2$. For main-sequence O stars in the canonical
mass range 20-60\,$M_{\odot}$, pressure scale heights are within the range
0.04-0.24 $R_{\odot}$, corresponding to a total number of clumps
6 $\times 10^3-6 \times 10^4$. These estimates
may in principle be tested through
linear polarisation variability, which probes wind asphericity
at the wind base.

In an investigation of WR linear polarisation variability 
Robert et al. (1989) uncovered an 
anti-correlation between the wind terminal 
velocity and the scatter in polarisation. They  
interpreted this as the result of blobs that grow or survive 
more effectively in slow winds than fast winds. 
Davies et al. (2005) found this trend to
continue into the regime of LBVs, with even lower $v_{\infty}$. LBVs are 
are thus an ideal test-bed for constraining clump properties, due
to the longer wind-flow times.
As Davies et al. found the polarisation angles of LBVs to vary 
irregularly with time, the line polarisation effects were attributed to 
wind clumping. 
Monte Carlo models for scattering off wind clumps have 
been developed by Code \& Whitney (1995); Rodriguez \& Magalhaes (2000); and Harries (2000), 
whilst analytic models to produce the variability of the linear polarisation 
may be found in Davies et al. (2007) and Li et al. (2009). 
Given the relatively short
timescale of the observed polarisation variability, Davies et al. (2007) argued 
that LBV winds consist of order 
thousands of clumps near the photosphere. 

For main-sequence O stars the derivation of 
wind-clump sizes from polarimetry 
has not yet been feasible as very high signal-to-noise data 
are required. This might become feasible with {\sc polstar}, {\sc Arago}, {\sc pollux-luvoir},
or other future space-based polarimeters. 
For now, LBVs provide the most promising types of test-objects owing to the 
combination of higher mass-loss rates, and lower terminal wind velocities. 

Linking back to theory, one of the key implications of
the LDI is that in multi-D hydrodynamical simulations the time-averaged $\dot{M}$ is {\it not}
anticipated to be affected by wind clumping, as
it has the same average $\dot{M}$
as the smooth CAK solution (Owocki et al. 1988). However, the shocked velocity structure and its
associated
density structure are expected to result in effects on the mass-loss
diagnostics. 

In contrast to the LDI simulations,
Muijres et al. (2011) studied the effects of clumping
on $a_{\rm rad}$ due to changes
of the ionisation structure, as well as the effects of wind porosity, using
Monte Carlo simulations. 
When only accounting for optically thin (micro) clumping $a_{\rm rad}$ was found to {\it increase} for
certain clumping stratifications $D(r)$, but only
for an extremely high clumping factor of $D\sim100$.
The reason $a_{\rm rad}$ may increase is the result of 
recombination yielding more flux-weighted opacity
from lower Fe ionisation stages (similar to the bi-stability physics).
For $D=10$ the effects were however found to be relatively minor.
When simultaneously also accounting for optically thick
(macro) clumping, the effects were partially reversed, as photons could 
now escape in between the clumps without interaction, and the predicted
$a_{\rm rad}$ goes down (see Muijres et al. 2011 for a range of 
clumping stratifications). Nevertheless, again,
for $D =10$ the effects were found to be rather modest.

The impact of wind-clumping on the predicted properties is presently still under debate and a fully consistent study has yet to be performed (see also Sundqvist et al. 2014).

\subsection{Empirical wind results in comparison to Theory} 
\label{sec_res}

Ideally this would be that part of the Review where all empirically determined mass-loss rates are collected and compared to theoretical predictions. 
Unfortunately, this is not yet feasible due to the limited amount of consistent observational analyses currently available. 
The largest homogeneous mass-loss samples for O-type stars are those of the VLT Flames surveys for low-$Z$ galaxies. 
However, in these studies only \ha\ was available to determine the mass-loss rate, leading to results which are clumping dependent. 
Using the wind-momentum luminosity relationship for the Galaxy, the LMC and the SMC, Mokiem et al. (2007) found 
\mdot\ vs. $Z^{0.83 \pm 0.16}$, in good agreement with theoretical predictions of Vink et al. (2001) and Bj\"orklund et al. (2021). They also noted reasonably good agreement with absolute mass-loss rates of Vink et al. (2001) in case the clumping factor is moderate ($D \sim$ 6-8; Ramirez-Agudelo et al. 2017). 

Bj\"orklund et al. (2021) on the other hand compared their predicted mass-loss rates to a small sample of clumping-corrected mass-loss rates from a variety of X-ray, UV, Optical and NIR data, both including and excluding porosity, finding good overall agreement. Future large and homogeneous analysis with clumping and porosity are needed to distinguish between different theoretical predictions. 

Ramachandran et al. (2019) found a notably steeper dependence of \mdot\ vs. $Z^{2}$, but their sample mostly concerned objects with intrinsically weaker winds. In other words, the empirical mass-loss $Z$ dependence may not be constant for all -- e.g. luminosity -- parameter regimes. 
Furthermore, Tramper et al. (2014) found larger than expected mass-loss rates in the even metal-poorer 
galaxies IC 1613, WLM, and NGC 3109, that have oxygen (O) abundances that are about 10\% solar. 
This mismatch of H$\alpha$ empirical mass-loss rates in these very low $Z$ galaxies by Tramper et al. and the radiation-driven
wind theory has been challenged by Bouret et al. (2015). So an in-depth analysis of all low-$Z$ O-type stars
with upcoming ULLYSES and complementary optical X-Shooter spectra is expected to shed more light on the $\dot{M}$ vs. $Z$ dependence.

Regarding terminal wind velocities, most studies overpredict them by about 25-40\% (Pauldrach et al. 1986; Muijres et al. 2012; Bj\"orklund et al. 2021). 
The predicted Vink \& Sander (2021) dependence of wind velocity on metallicity as $v_{\infty} \propto Z^{0.19}$ for 
O stars appears to agree with observations (Garcia et al. 2014), but it too early to confirm convincingly. Bj\"orklund et al. (2021) provide a relation in the opposite direction $v_{\infty} \propto Z^{-0.10}$, highlighting theoretical complexities. Both dependencies are weak, and larger data-sets are warranted for discrimination.
Garcia et al. (2014) have shown the complications of their $Z$ dependence from the empirical side, and also shown that the
Fe abundance in these galaxies is larger than expected on the basis of nebular O, and
more similar to the Fe abundance of the SMC.
In other words, empirical studies appear to show a sub-solar [$\alpha$/Fe] ratio.
Non-solar metallicity scaled [$\alpha$/Fe] conversions as a function of metallicity are given in Table 5 of Vink et al. (2001).

With respect to the \teff\ dependence below the BS-Jump, the Vink \& Sander (2021) drop of $v_{\infty}$ towards lower \teff\ appears to be in line with observations, but the predicted mass-loss jump (Vink et al. 1999; Krticka et al. 2021) is still elusive (Markova \& Puls 2008). As was discussed in Sect.\,\ref{sec_bow}, the recently developed bow-shock method might shed light on the issue of the very existence of the BS-Jump. While the mass-loss rates derived by Kobulnicky et al. (2019) might have systematic uncertainties\footnote{The absolute rates are lower than the Monte Carlo predictions of Vink et al.}, in a relative sense there is no reason why this method would systematically treat objects on the hot or cool side of the BS-Jump differently. In other words, the Kobulnicky et al. finding of a mass-loss increase by a factor of a few at the BS-Jump (see their Fig.\,9) could be highly informative.

Another area where this model-independent mass-loss rate would help is in the regime of the "weak-winds".
Vink \& Sander (2021) uncovered capricious behaviour in the wind properties of locally consistent wind models 
around 35\,kK, which might be linked to earlier issues that Lucy (2010) and Muijres et al. (2012) reported for lower luminosity models, and where this was attributed
to the onset of the "weak-wind problem" (Martins et al. 2005; Puls et al. 2008; de Almeida et al. 2019). 
This issue might be related to the `inverse' bi-stability effect
due to a lack of Fe {\sc iv} driving around spectral type O6.5 -- corresponding to $\log(L/\lsun) = 5.2$.
If the `weak wind problem' is not only directly related to low L, but also the specific Teff of approximately 35\,kK we
may possibly expect objects of order 40\,kK and higher to again show stronger wind features, even when the objects remain below 
$\log(L/\lsun) = 5.2$. 
This should be testable with the new ULLYSES observations for O-type stars in this temperature range.
It may also be feasible to be able to witness extremely fast winds with $v_{\infty} > 5\,000$ km/s for the weak wind stars around 35\,000 K.

The presence or absence of such a regime with weaker, faster winds in large observational samples such as X-Shooting ULLYSES (XSHOOTU) will mark an indicator for all current mass-loss prediction efforts that could provide important hints as to whether or not additional driving physics needs to be considered to explain empirical findings.
Ultimately, only when radiative pressure computations are directly combined with empirical mass-loss rates that include consistent velocity stratifications (Sander et al. 2017) will it become clear if all relevant physics is included in state-of-the-art atmosphere models.

For evolved WR stars the most common methods have been the radio method (e.g. Nugis \& Lamers 2000) and the 
Non-LTE model atmosphere emission-line method (e.g. Hamann et al. 1995). Both of these methods thus rely on density squared diagnostics and are hence dependent on the clumping factor $D$. 
While for most O-type stars the factor $D$ is difficult to constrain
from observations in a model-independent manner, the clumping factor can
be obtained in a relatively accurate manner in WR winds thanks to the
electron scattering wings of strong emission lines (Hillier 1991), which are linearly dependent on the density. Typical derived
values are $D = 4 \dots 16$, i.e. mass-loss reductions of factors 2-4 in
comparison to non-clumped analyses, and in agreement with assessments of
Moffat \& Robert (1994).
The Hamann et al. (1995) analyses have been updated with line-blanketed analyses (Gr\"afener et al. 2002; Crowther et al. 2002) and more recently for WN stars in the Galaxy with Gaia distances (Hamann et al. 2019) to complement analyses in the LMC (Hainich et al. 2014) and the SMC (Hainich et al. 2015; Shenar et al. 2016), while a new Gaia study of Sander et al. (2019) attacked the Galactic WC stars. The smaller subset of WO stars at lower $Z$ was studied in Tramper et al. (2015). What these analysis have in common is a more or less model independent clumping factor of approximately $D = 10$ (from electron-scattering wings), and a general existence of a host galaxy $Z$ dependence, but the appropriate comparisons of empirical rates with theory still need to be performed, as theoretical relations are only just starting to appear.

\begin{summary}[SUMMARY POINTS]
\begin{enumerate}
\item Several methodologies to derive mass-loss rates have been discussed. These involve traditional diagnostics, such as H$\alpha$, UV P Cygni lines, and radio free-free emission, as well as newer ones including X-rays and the bow-shock method. 
The different methods still need to be combined for large samples, such as XSHOOTU/ULLYSES.
\item Wind clumping has been recognised as a key feature effecting empirical mass-loss rates, but different diagnostics, such as H$\alpha$ and the UV lines react in different ways to optically thin and optically thick clumps. 
\item The origin of wind clumping, whether sub-photospheric and/or resulting from wind instabilities such as the LDI still needs to be addressed. Only when this challenging task has been successfully undertaken can we be confident that the wind clumping implementations into model atmospheres are physically sound, and mass-loss rates can be considered accurate as input for stellar evolution modelling.
\end{enumerate}
\end{summary}

\section{Stellar Evolution with Mass Loss}
\label{sec_evol}

Over the last five decades many researchers have studied one particular effect of mass loss on stellar evolution models, involving the loss of the envelope mass due to stellar winds (the "Conti" scenario). This is what could produce WR stars (with binary RLOF being an alternative) and sets the black hole mass function with metallicity $Z$. A second effect is that of the loss of angular momentum (e.g. Langer 1998; Maeder \& Meynet 2000) discussed in Sect.\,\ref{sec_rot}. A more recent development is that of wind-envelope interaction (Grassitelli et al. 2021) which shows that in order to understand LBV S\,Dor variations it is critical to study the effects of envelope mass loss on the actual structure of the outer envelope (Sect.\,\ref{sec_struc}) rather than simply loosing an "x" amount of mass from the outer layers as has been considered up to now. 
It should be realised that the single star physics of wind-envelope interaction is equally relevant for single star as binary evolution.

\subsection{Loss of Envelope mass}
\label{sec_mloss}

\begin{figure}
\includegraphics[width=1.0\columnwidth]{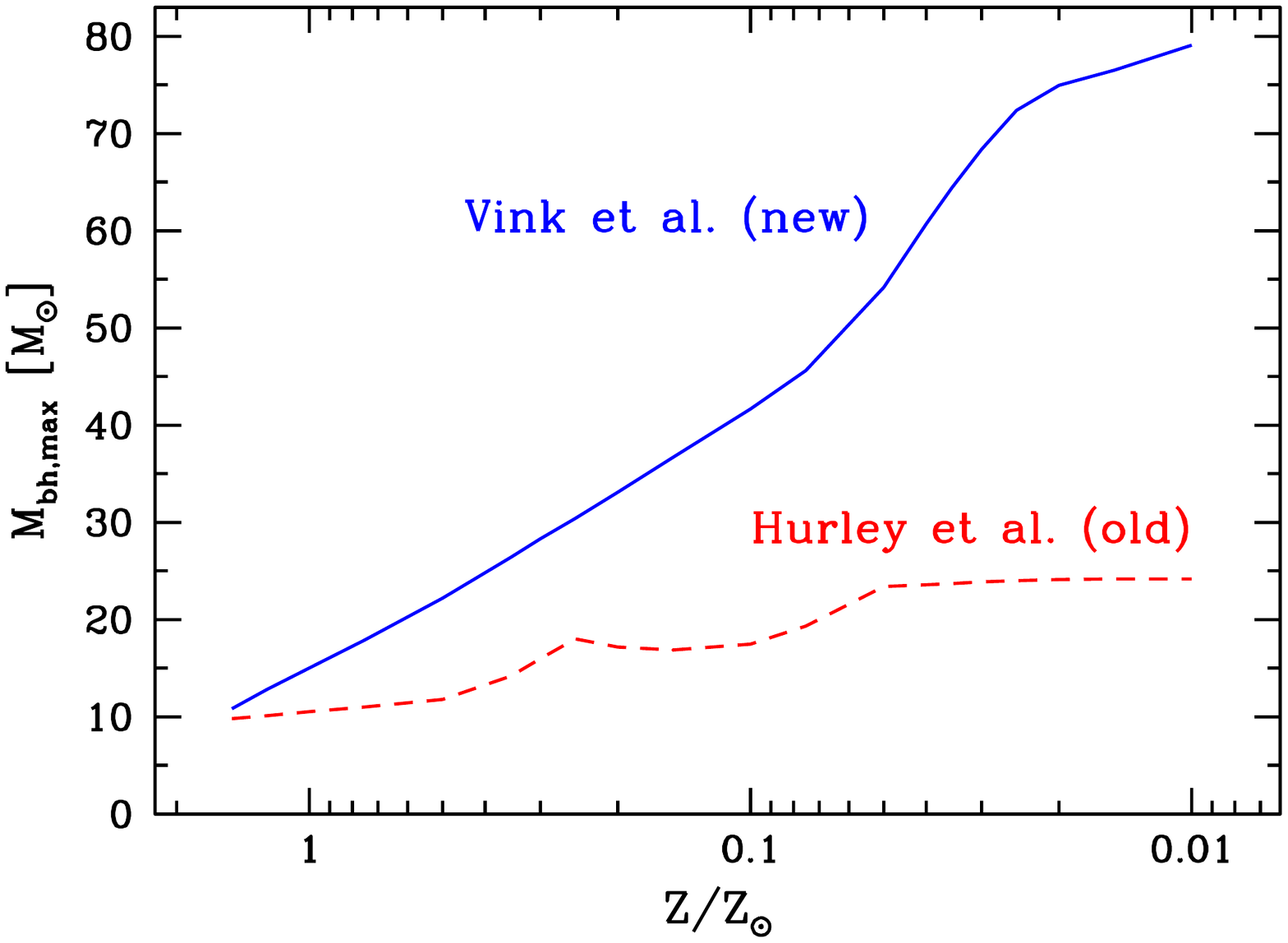}
\caption{The dependence of the Maximum black hole mass below the pair-instability gap on $Z$ for previous sets of stellar winds from Hurley et al. and for 
$Z$-dependent  
Vink et al. winds. Note that the maximum BH mass is similar for solar-like
metallicities, while it is significantly larger at lower $Z$ when Fe-dependent WR winds are accounted for (Vink \& de Koter 2005; Eldridge \& Vink 2006). The figure is adopted from Belczynski et al. (2010), while a similar plot was also employed in the analysis for a low-$Z$ progenitor of the first gravitational wave event GW\,150914 (Abbott et al. 2016).}
\label{f_bhmax}
\end{figure}

The most dramatic effects of mass loss might be seen for very massive stars, and as the H-burning phase constitutes 90\% of its life-time, the initial phases are the most important (see Fig.\,\ref{f_evap}). For more canonical stars around 60\,$M_{\odot}$ stellar mass is lost continuously during all phases of massive star evolution. Groh et al. (2014) nicely illustrates how this occurs during the H-burning main sequence, the He burning BSG phase as well as the final WR stage -- in roughly equal quantities. It is thus important to figure out how robust the absolute amount of mass loss is during all \teff\ phases, as well as its $Z$ dependence. 
The relevant aspect for the BH mass function is that the mass-loss rate depends on Fe 
(See Figs.\,\ref{f_mdotz2} and \ref{f_mdotWR}). It was this $Z$-dependence that is the key factor for predicting larger BH masses in lower $Z$ environments, such as for GW\,150914 (Abbott et al. 2016) as shown in Figure\,\ref{f_bhmax}. 
Note that heavy black holes are only possible for Galactic $Z$ if the winds are somehow quenched by a strong magnetic field (Petit et al. 2017).

The $Z$ dependence is not only critical for 1D models of stellar envelope mass, but when occurring in conjunction with rotation, the loss of angular momentum becomes highly relevant as well, especially in the context of long GRBs.

\subsection{Angular momentum loss}
\label{sec_rot}

Massive stars rotate at significant speeds and this may directly affect both the stellar structure (see Gagnier et al. 2019 for 2D stellar evolution modelling) as well as the geometry and strength of stellar winds. The first potential effect is that of the centrifugal force leading to a lower effective escape speed and higher mass-flux from the stellar equator (Friend \& Abbott 1986; Pauldrach et al. 1986). 
However, due to the physics of Von Zeipel 's gravity darkening the radiative flux from the pole is higher than that from the equator, leading to a polar wind for extremely luminous stars (Owocki et al. 1996) such as $\eta$ Car. Wind compression effects in Be stars (Bj\"{o}rkman \& Cassinelli 1993) and WR stars (Ignace et al. 1996) may play complementary roles.

For most "canonical" O-type stars the various rotational effects are probably relatively modest (see Petrenz \& Puls 2000). This is consistent with results from linear spectropolarimetry studies of O-type stars that show that most garden variety O-stars are unpolarised (Harries et al. 2002), and while many 1D stellar evolution calculations have included enhanced mass-loss rates there is surprisingly little evidence that this is actually correct (M\"uller \& Vink 2014). 

Related to rotation, there is also the issue of dipolar fields that might brake the magnetic subset of O-stars (ud-Doula et al. 2002).

\subsection{Mass-Luminosity plane}
\label{ML-plane}

\begin{figure}
\centering
\includegraphics[width = 8cm]{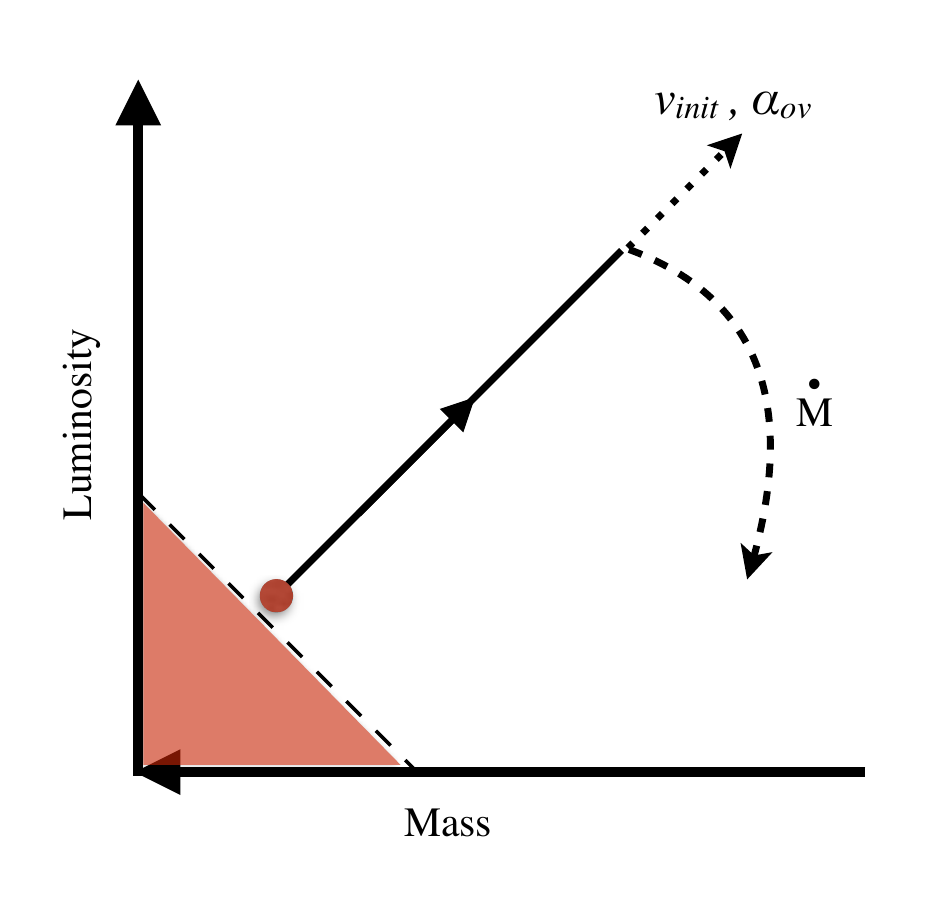}
\caption{Illustration of the Mass-Luminosity plane with a typical evolutionary path entering the ZAMS at the red dot, evolving along the black arrow. The dotted vector indicates how increased mixing may extend the $M-L$ vector. The curved dashed line represents the gradient of how mass loss affects this $M-L$ vector. The red solid region provides the boundary set by the mass-luminosity relationship, forming a forbidden zone (Higgins \& Vink 2019).}
\label{f_MLplane}
\end{figure}

Until recently, mass-loss rates included in stellar models had only been constrained through stellar wind density studies, which unfortunately depend on uncertain clumping properties. 
However, it has now also become possible, at least for the main-sequence phase, to constrain $\dot{M}$ in an independent way, through the consideration of vectors in the Mass-luminosity plane (Higgins \& Vink 2019). Figure \ref{f_MLplane} shows a typical evolutionary path of a star that evolves to lower \teff, indicating two 
key physical processes in massive star evolution: mixing and mass loss. These uncertain processes follow very different vectors in the ML-plane.
Mixing, which can involve core overshooting, rotational mixing, or any other -- possibly even yet to be identified -- process follows the vector that extends the path of the star to lower \teff\ and higher $L$. 
Stronger $\dot{M}$ on the other hand leads to a drop in $L$ and $\dot{M}$ is constrained by the {\it gradient} (see the curved dashed line). 

For the eclipsing 40 $+$ 30 $M_{\odot}$ binary HD\,166731 (Mahy et al. 2017), Higgins \& Vink (2019) were able to constrain the mass-loss rates within about a factor of two of the Vink et al. recipe, while they simultaneously presented evidence for enhanced core overshooting in the range $\alpha_{\rm ov} = 0.3 -0.5$ for both components of the binary system. 
Core overshooting sets the timescale and hence the total amount of mass loss during evolution on the main sequence. It thus depends on $\dot{M}$ vs. \teff. Overshooting remains a key uncertainty in massive star evolution modelling and may be constrained from asteroseismology (e.g. Aerts et al. 2019; Johnston 2021). Bowman (2020) provides a Table of $\alpha_{\rm ov}$ for stars up to $\sim$25\,$M_{\odot}$ from asteroseismology studies with values as low as 0 and as high as 0.44. 
The adopted values are very different for different evolutionary models. For instance the Bonn models employ a relatively large value of 0.335 (Brott et al. 2011), while the Geneva models use 0.1 (Ekstr\"om et al. 2012). 
The adopted values are crucial when computing evolutionary paths for the most massive BH progenitors such as the one recently found in the "pair instability" upper BH mass gap of 85\,$M_{\odot}$ in gravitational wave event GW\,190521 (Vink et al. 2021; Tanikawa et al. 2021).

\subsection{Wind-envelope interaction}
\label{sec_struc}

\begin{figure}
\begin{center}
\includegraphics
  [width=\textwidth,angle=90]{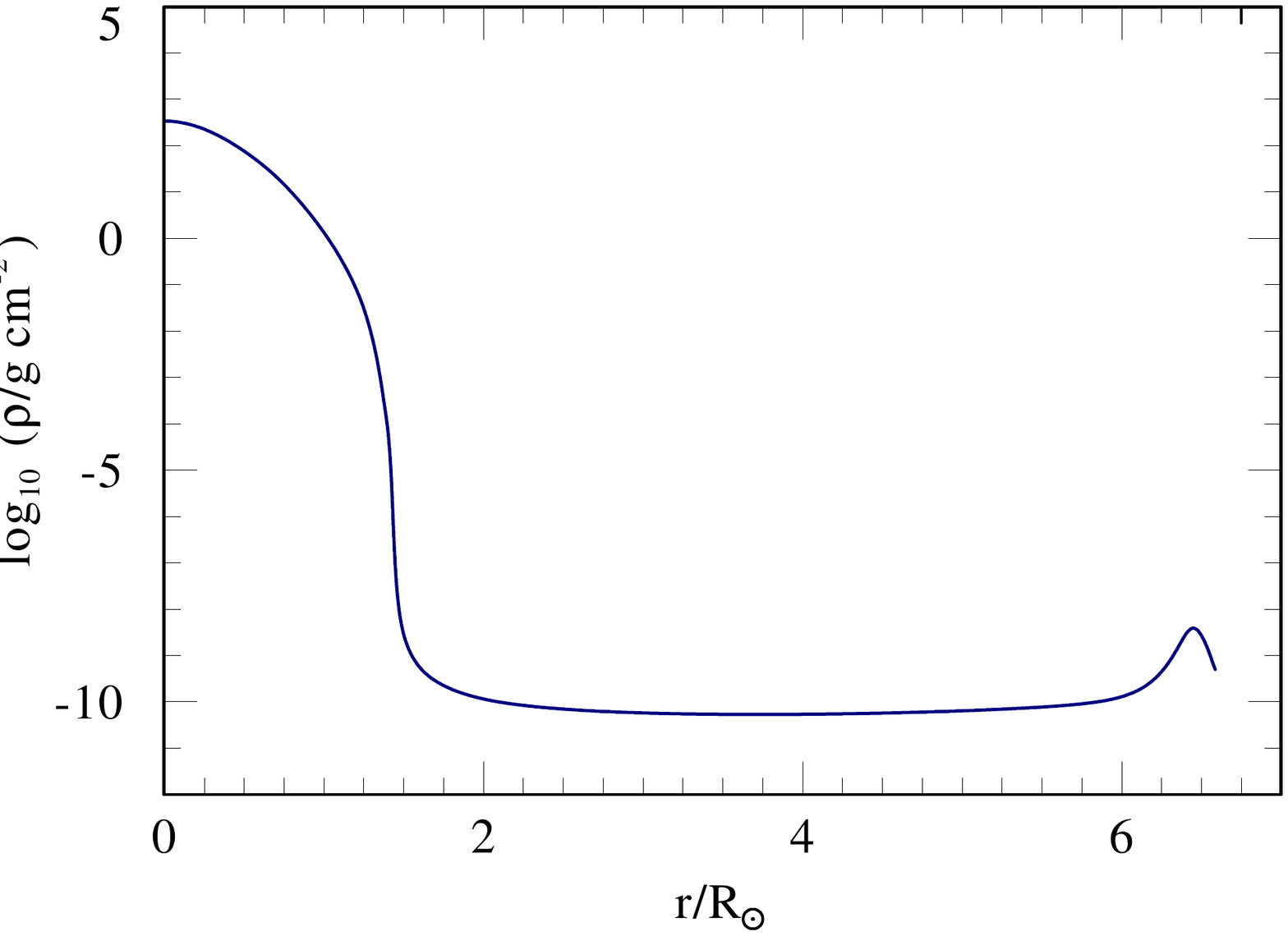}
\caption{The peculiar density structure of a massive star with an inflated envelope. The model
has a very extended low-density envelope over many stellar radii (Gr\"afener et al. 2012). This structure is notably different from a star that simply expands red-wards after H exhaustion on the main sequence. Inflated envelopes are specific to radiation-pressure dominated envelopes in close proximity to the Eddington limit.}
\label{f_infl}
\end{center}
\end{figure}

The basic physics of radiation-dominated envelopes of massive stars in close proximity to the Eddington limit is very uncertain. When searching for a hydrostatic solution, the Fe-bump and other opacity bumps in OPAL opacity tables would lead to the "inflation" of the outer stellar layers (Ishii et al. 1999; Gr\"afener et al. 2012; Sanyal et al. 2015) . Note that 
inflation is notably different from canonical envelope expansion after core H exhaustion. Figure\,\ref{f_infl} shows the peculiar density structure of such an inflated static 1D stellar structure model from Gr\"afener et al. (2012), where it was suggested how inflated envelopes may play a key role in the S\,Dor variations of LBVs. More recently, Grassitelli et al. (2018; 2021) updated the Bonn stellar evolution code BEC with a hydro-dynamic outer boundary condition. For LBVs characteristic of an object such as AG\,Car a 
gap in the gradient of OPAL opacities was identified between 20-30 kK. 
If this "forbidden" temperature region were subjected to an increasing mass-loss rate with lowering \teff\ as predicted by the BS-Jump, Grassitelli et al. (2021) predicted "cycles" of hotter and cooler envelope configurations, entertaining the first ab-initio stellar evolution prediction of the S\,Dor LBV 
instability on the correct timescale (see Humphreys \& Davidson 1994 for previous suggestions). 

\begin{marginnote}
\entry{S Doradus Variables}{The most characterising variability of LBVs. Named after the LBV prototype S Dor located in the LMC. Famous Galactic Members include AG Car and HR Car (see Richardson \& Mehner 2018 for a recent inventory). }
\end{marginnote}

A key uncertainty concerns the mode of energy transport (convection or radiation) in these inflated envelopes. Some stellar evolution codes remove 
the inflation (e.g. Yusof et al. 2013), possibly even somewhat artificially. Another technique involves that of the usage of the MLT$++$ routine in MESA. 
3D radiation hydrodynamics computations by Jiang et al. (2015) show the formation of a clumpy, porous medium, which may affect energy transport when 
incorporated in 1D stellar models. 

\begin{summary}[SUMMARY POINTS]
\begin{enumerate}
\item Evolutionary modelling is not only affected by our knowledge of absolute mass-loss rates, but also by the manner in which the rates vary over time. It is therefore important to construct correct $\mdot$
dependencies on the stellar parameters.
\item Internal mixing such as overshooting affects these stellar parameters, as well as stellar lifetimes, and thus the overall amount of mass lost. The problem of stellar evolution with mass loss is that the construction of the mass-loss history is non-linear.
\item It is not yet established how additional physical effects such as rotation and binary affect the winds and evolution. On the one hand, winds are able to affect the rotational evolution, and on the other hand stellar rotation might affect the wind strength and latitudinal dependence, both of which are not yet settled. This requires observational guidance from polarimetry observations. 
\item The envelope structures of massive stars in close proximity to the Eddington are very poorly understood, but they are directly affected by stellar winds. Despite recent progress in wind-envelope interaction of LBV S\,Doradus behaviour, there are still many questions about the correct inclusion of the LBV phase in stellar models.
\end{enumerate}
\end{summary}

\section{Variable and Pre-supernova mass loss}
\label{sec_lbv}

So far, we have focussed our attention on stationary stellar winds during the stable H and He phases of stellar evolution and nucleosynthesis. 
As these phases dominate the lifetimes they are probably the main contributors to the overall mass-loss budget during massive star evolution. 
However, we know of cases -- such as $\eta$ Car -- where up to 10 solar masses of material was expelled over a period of just a few decades. 
If such events were common, these type of Giant Eruptions (SN-Imposters) could contribute significantly to, and at lower $Z$ even dominate, the overall 
mass-loss budget (see Smith 2014). For these reasons it is important to find out how common LBVs are, what sets the physics of their eruptions, and 
what is the duty cycle (Massey et al. 2007; Kalari et al. 2018). 
For instance, the only other Galactic Giant Eruptor P\,Cygni has a nebular mass of just 0.1$M_{\odot}$, 
i.e. a factor of 100 lower than for $\eta$ Car. If the P\,Cygni result is more typical than that of $\eta$ Car this would imply that eruptive mass loss would be almost irrelevant for the total mass-loss budget during stellar evolution, unless the LBV phenomenon occurs hundreds of times. 
It is hence pivotal to understand the physics of the LBV phenomenon. 
Prior to jumping to the most extreme cases like $\eta$ Car, there is a need to understand the common S Dor mass-loss cycles, necessary in order to get a realistic impression of the impact of the LBV phenomenon in general.

In addition to the LBV phenomenon, the terms "episodic" and "eruptive" mass loss have recently also been used in SN literature, albeit often with a very different meaning.
Let us take a moment to clarify some relevant terminology. In recent SN literature terms 
such as episodic and eruptive, as well as high Eddington, superwind, super-Eddington have oftentimes been used intermittently leading to some confusion.
Let us first address the issue of episodic mass loss. In some cases, gigantic mass loss such as that of $\eta$ Car could be both episodic and eruptive. 
However, generally just because mass loss is episodic this would not automatically imply that it has an eruptive nature. S\,Dor variations for instance exhibit episodic mass loss but their behaviour appears to be well explained by (quasi)stationary winds, without the need to invoke eruptions.

In line with its use by the Asymptotic Giant Branch community, the term superwind describes the cumulative
effect of increasing wind mass-loss rates when stars become more luminous. These winds could be radiation-driven and of increasing strength, for instance when the Eddington factor becomes of order $\sim$0.5 (Gr\"afener \& Vink 2016). 
Alternatively, RSGs in close proximity to the Eddington limit could exhibit pulsational mass loss (Heger et al. 1997; Yoon \& Cantiello 2010).
Following Moriya et al. (2020) in the context of the intriguing iPTF14hls a hyperwind has an extremely high mass-loss rate more than a few $M_{\odot}$ per year,
which cannot be explained by canonical RDWT. Here, the physics of continuum-driven super-Eddington winds may become relevant, as discussed below.

\subsection{Luminous Blue Variable winds}

Although most S\,Dor variables have been subject to photometric
monitoring, only a few have been analysed in sufficient
detail to understand the driving mechanism of their winds.  
Mass-loss rates are of the order of $10^{-3}$ - $10^{-5}$ \msunyr,
whilst terminal wind velocities are in the range $\sim100-500$\,km/s. Obviously, 
these values vary with $L$ and $M$, but there are indications
that the mass loss varies as a function of $\Teff$ when the S\,Dor variables
transit the upper HRD on timescales of years. As the stellar luminosity remains roughly constant during S\,Dor variations, they provide an ideal laboratory for testing the RDWT as a function of \teff\ for individual objects.

The Galactic LBV AG\,Car is one of the best studied S\,Dor variables. 
Vink \& de Koter (2002) predicted $\dot{M}$ varies in line with
radiation-driven wind models for which the $\dot{M}$ variations 
are attributable to ionisation shifts of Fe. Sophisticated non-LTE spectral
analyses have since confirmed these predictions (Groh et al. 2009). 
It is relevant to mention here that this variable wind concept 
(bi-stability) has also been suggested to be responsible for CSM 
density variations inferred from 
modulations in radio light-curves and H$\alpha$ spectra of Type IIb and IIn SNe (Kotak \& Vink 2006; Trundle et al. 2008). 
However most stellar evolution models would have predicted massive stars
with $M$ $\ge$ 25\,$M_{\odot}$ to explode at the end of the WR phase, rather than
after the LBV phase. These implications could be gigantic, 
impacting our most basic understanding of massive star death (Gal-Yam \& Leonard 2009; Smith 2014; Justham et al. 2014).

\subsection{Super-Eddington winds}

Whilst during ``quiet'' S\,Dor phases, LBVs may lose significant amounts of mass 
via ordinary line-driving, some objects are subjected to phases of extreme mass loss. 
For instance, the giant eruption of $\eta$ Car with a
cumulative loss of $\sim$10\,$M_{\odot}$ between 1840 and 1860 (Smith et al. 2003)
which resulted in the famous Homunculus nebula corresponds to an effective mass-loss rate 
a factor of 1000 larger than that expected from the RDWT for an object of that luminosity. 

Owocki et al. (2004) studied the theory of
porosity-moderated continuum driving for objects that would be considered to  
exceed the Eddington limit. 
Continuum-driven winds in super-Eddington stars could reach mass-loss rates approaching 
the {\it photon tiring limit}, $\mdot_{\rm tir} = L/(GM/R)$, which should result in a stagnating 
flow that would lead to spatial structure in multi-D simulations (van Marle et al. 2009). 
However, alternatively clumping may result from instabilities relating to 
Fe opacity peaks (Cantiello et al. 2009; Gr\"afener et al. 2012; Jiang et al. 2015; Grassitelli et al. 2021; 
Glatzel et al. 1993), especially for objects approaching 
the $\Gamma$-limit. 

The equation of motion can be approximated as: 

\begin{equation}
v\Bigl(1-\frac{a^2}{v^2}\Bigr)\frac{{\rm d}v}{{\rm d}r} \simeq
g_{\rm grav}(r) + a_{\rm rad}(r) =-\frac{GM}{r^2}(1-\Gamma(r)) .
\end{equation}
At the sonic point, $r_{\rm s}$: $v=a$, and thus 
$a_{\rm rad} = -g_{\rm grav}$ implying $\Gamma(r_{\rm s})=1$.
In order to have a wind solution we require $\Gamma$ to cross unity at $r_{\rm s}$.
In other words, $\Gamma(r)$ must be $< 1$ below the sonic point 
and $\Gamma(r)$ must be $>1$ above it. 
An accelerating wind solution thus implies an increasing opacity$\frac{{\rm d}\bar \kappa}{{\rm d}r}|_s>0$. 

Without sub-sonic opacity reduction due to porosity, one would not have the situation that $\Gamma$ crosses unity at the sonic point, and the entire atmosphere would be Super-Eddington with $\Gamma(r) >1$.
However, {\it if} the sub-sonic opacity could be reduced, e.g. due to the formation of a porous medium (Shaviv 2000) -- which is quite possible near the Eddington limit -- a continuum-driven wind solution becomes viable. 
The reason is that outward travelling photons could avoid the optically thick clumps, lowering $a_{\rm rad}$ 
below the sonic point, and the effective Eddington parameter might drop below unity. 
Further out in the wind, the clumps become optically thinner -- as a result of expansion -- and the porosity effect thus less relevant, and $\Gamma$ could be larger than 1. 
In other words, a wind 
solution with $\Gamma$ crossing unity is feasible, even for 
objects that are formally {\it above} the Eddington limit.

Owocki et al. (2004) expressed the effective opacity in terms 
of the porosity length, showing 
that the mass-loss rate might 
become substantial when the porosity length becomes of the order of the 
pressure scale height.  They developed a concept
of a power-law distributed porosity length (in analogy to the CAK-type 
line-strength distribution), and showed 
that even the gigantic mass-loss rate during Eta Car's 
giant eruption of 10$M_{\odot}$ might be explicable by some form of 
radiative driving. 
In more recent times, Moriya et al. (2020) invoked continuum-driven winds as a possible explanation for a so-called "hyperwind" in iPTF14hls.

\subsection{CSM interaction}

\subsubsection{Type IIn} 

Both (quasi)stationary stellar winds and eruptive events can lead to regions of high-density circumstellar material (CSM) in the close surroundings of a massive star that is about to expire. If the star explodes as a core collapse SN event the ejecta ram through the CSM leading to both a forward and backwards shock. The photons produced in the shock ionise the un-shocked CSM in front of the SN ejecta (e.g. Chevalier \& Fransson 1994; Chugai et al. 2004). This matter subsequently recombines giving rise to narrow spectral lines that reflect the low velocities of the prior wind or eruption. This has been the basic picture for the common interacting Type IIn (e.g. Fillipenko 1997) class. Physical mechanisms for the large amount of gaseous material inferred from the dense CSM are pulsational pair instability mass loss from VMS (Woosley \& Heger 2015) or gravity waves that could potentially unbind a few $M_{\odot}$ in a few months to decades before demise (Quataert \& Shiode 2014; Wu \& Fuller 2021).

There are also cases, such as SN 2009ip (Mauerhan et al. 2013; Fraser et al. 2013; Ofek et al. 2014) 
where the nature of the final explosion is still being debated. Intriguing light-curves show peaks prior to the final peak, and it is thus unclear if the star survived the eruption or not. If it has survived it is not a IIn SN but a giant eruption of an LBV like star: a SN imposter. 

As a word of caution I note that even in those extreme cases where VMS show evidence for solar-masses type eruptions in just a few years prior to explosion, this would not at all imply that eruptive mass loss dominates the mass-loss budget accumulated during stellar evolution. 
On the contrary it is probably even the other way around: the most massive stars have the strongest mass-loss rates that may evaporate already on the main sequence (recall Fig.\,\ref{f_evap}).

\subsubsection{Flash Spectroscopy}

Over the last number of years, SN explosions in dense CSM have been probed more regularly 
thanks to more efficient response facilities for transients. 
Gal-Yam et al. (2014) discovered the emission-line progenitor with a strong ($10^{-3}$ \msunyr) stellar wind in the type IIb SN 2013cu taken just 15 hours after explosion. This new spectroscopic method provides fantastic opportunities for stellar wind and atmosphere modelling to play a direct role in the identification of interacting SNe. The first case of 2013cu involved a transitional SN with only a small amount of H left over. 
The progenitor was likely an LBV or YSG (Groh 2014). Subsequent analyses including light-travel time effects showed that the He {\sc I} line formed at a large radius indicating an extended CSM and a canonical wind that had been blowing for at least 70 years, and possibly longer (Gr\"afener \& Vink 2016). 
This illustrates that at least some massive stars may be undergoing a superwind phase and CSM interaction without necessarily having undergone Giant eruptions of several solar masses. 
In more recent studies it appears that even in the 
most common type IIP SNe -- which at first sight do not show evidence for a CSM  -- in fact do show interaction features 
at early times, and it will be interesting to find out if stellar superwinds or eruptive mass loss events are the dominant aspect of such pre-SN mass loss.

\subsubsection{Type I}

In addition to Type IIn and transitional IIb cases, there are Type Ibn supernova events. The first such case was SN 
2006jc (Pastorello et al. 2007) where the narrow lines highlight that strong pre-SN mass loss is not confined to H-rich cases. Whether these events are due to WR type single stars or binary interactions is still under debate (Sun et al. 2020) but what is clear is that interaction takes place in the H-poor Universe as well.

This is particularly relevant in the case Superluminous Type I supernovae. These SN might possibly become key cosmic distance indicators (e.g Inserra et al. 2021) but the origin of their extra brightness is still under debate. Given the similarity to the Type IIn SNe the most obvious explanation is that of CSM interaction, as even two shells without invoking an explosion may produce huge amounts of radiation (Dessart et al. 2009). Magnetar powering however is a viable alternative (Kasen \& Bildsten 2010). In some cases the light-curve might be so long-lived from nickel decay that PISNs (Barkat et al. 1967) are the third and final option for at least a subset of SLSNe. 

The evolution of these VMS into the PISN regime is obviously heavily dominated by mass loss, so the PISN question and $Z$-dependent mass loss physics are intimately related (e.g. Langer et al. 2007: Yusof et al. 2013; Yoshida et al. 2016).

\begin{summary}[SUMMARY POINTS]
\begin{enumerate}
\item While the S Dor phase of LBVs are starting to become understood in terms of the RDWT as a function of effective temperature and envelope modelling, super-Eddington giant LBV Eruptions, such as $\eta$ Car and SN Imposters remain a mystery.
\item The CSM can be probed to understand pre-SN mass loss and help identify SN progenitors. Traditionally this has been performed by multi-wavelength methods, including X-rays and radio obsevations.
\item The novel method of Flash Spectroscopy can provide additional information on progenitor mass loss, as not only the density but also the wind velocity can be probed directly, helping identify SN progenitors.
\end{enumerate}
\end{summary}

\section{Summary and Future Directions}

As we have seen in this review, accurate mass-loss rates are important for a proper understanding of the stellar parameters of massive stars as the derived quantities depend on the hydro-dynamic wind structure $v(r)$, for which $a_{\rm rad}(r)$ needs to be determined. Furthermore stellar evolution modelling is heavily affected by assumed empirical or theoretical mass-loss recipes, both in terms of the loss of stellar mass and angular momentum. A more recent development is that the envelope structures depend strongly on the mass-loss behaviour with \teff, which affects both single star and binary evolution. 

Most stellar evolution models utilise the smooth 
Monte Carlo {\it theoretical} predictions of Vink et al. (2000).
However, it has become clear that {\it empirical}
$\dot{M}$ rates have been overestimated when determined from 
$\rho^2$ diagnostics such as \ha. For O supergiants 
unclumped \ha\ rates however are generally 
a factor 2-3 {\it higher} 
than the Vink et al. (2000) rates, meaning that 
moderate clumping effects ($D$ = 4-10) may 
indirectly already be accounted for in current stellar evolution models.
However, more recent analyses that account for clumping and porosity appear to provide $D$ factors 
larger than 10 which may be in better agreement with CMF computations such as those of Petrov et al. (2016), Sander et al. (2017), Krticka \& Kubat (2017) and  Bj\"orklund et al. (2021) that are typically a factor 2-3 lower than the Sobolev-based Monte Carlo predictions. 

Moreover, there has been progress in our qualitative and quantitative understanding 
of $\dot{M}$ for stars that approach the Eddington $\Gamma$ limit. 
Vink et al. (2011) discovered a ``kink'' in the $\dot{M}$ 
vs. $\Gamma$ relation at the transition from optically thin O-type
to optically thick winds, and this mass-loss enhancement is crucial for the evolutionary paths of the most massive stars.
I also discussed a methodology that involves a model-{\it independent} $\dot{M}$ 
indicator: the transition mass-loss 
rate $\dot{M}_{\rm trans}$ -- located exactly at the transition from optically thin to optically thick stellar winds. 
As $\dot{M}_{\rm trans}$ is model independent, {\it all} that is required 
is to determine
the spectroscopic transition point in a given data-set and to determine 
the far more accurate $L$ parameter.
In other words $\dot{M}_{\rm trans}$ is extremely useful for 
calibrating wind mass loss. The situation is that the transition mass-loss rate is in good agreement with empirical mass-loss rates with $D=10$ and Vink et al. (2000) predictions, and that an extrapolation of CMF based models such as Bj\"orklund et al. (2021) rates to higher mass yields mass-loss rates that are a factor 2-3 too low. Future CMF hydro computations are needed to resolve these issues. 

Furthermore, clumping may affect mass-loss predictions in various ways, and I have 
highlighted that both the origin and onset of wind clumping remain elusive. 
Polarisation measurements call for clumping to be already present 
in the stellar photosphere, but how this would interact with 
the hydro-dynamical LDI simulations further out, and how this would 
need to be consistently incorporated into 
radiative transfer calculations and mass-loss predictions is as yet 
unclear. For these reasons, the search for the 
nature and implications of wind clumping should certainly continue, and future
UV polarimeters such as {\sc polstar}, {\sc arago}, or {\sc pollux-luvoir} 
would provide marvellous opportunities.

Another key aspect is that of the \teff\ dependence of $\dot{M}$. Different mass-loss prescriptions give different qualitative behaviours, and it is critical that empirical studies 
would test this, especially with respect to the first and second BS-Jumps, as they have a large effect 
on stellar evolution as well as envelope structure modelling. Long timescale variability studies (months, years) including LBVs and other pre-SN mappings will become possible with new transient surveys such as LSST/Rubin, while shorter timescale variability facilities, such as TESS, may study stochastic low-frequency variability to test if the data are due to the LDI (Krticka \& Feldmeier 2021) or more consistent with either internal gravity waves (e.g. Bowman et al. 2019) or sub-surface convection (Cantiello et al. 2021).

While many multi-D effects are being studied, we are still quite far away from simultaneous 3D radiative transfer and hydrodynamics. As most massive stars are spherical on the larger scales this will probably not matter too much in terms of the global physics, but in terms of the micro-physics it is critical. State-of-the-art Multi-D hydro 
simulations (e.g. Sundqvist et al. 2018) are currently still performed using simplified parametrizations of the line acceleration, while on the other hand, state-of-the art stationary non-LTE CMF models are still 1D for a while to come, and dynamically consistent diagnostic modelling has only just become possible (Sander et al. 2017). Multi-D radiative transfer to test the wind driving physics are being developed (e.g. Hennicker et al. 2018) and this will be most helpful when better multi-D data in terms of interferometry and linear spectro-polarimetry become available in the future. 

For the moment we will still need to rely on phenomenological models that can implement clumping \& porosity in 1D
(Surlan et al. 2013; Sundqvist \& Puls 2018). Large homogeneous samples in both the optical (for \ha) and the UV are needed to make substantial progress on empirical mass-loss relations. This combined approach will be possible with the XSHOOTU project that combines ULLYSES UV data with optical ESO-VLT data for approx 250 massive stars in the LMC and SMC, while a pilot study of even lower $Z$ galaxies like Sext A is also included. These analyses involve time consuming non-LTE modelling processes and analysis tools, but we can be confident that we will soon have an empirical $\dot{M}$ versus $Z$ relation over a decent range of $Z$. 

This will be of paramount importance for the next step: the understanding of the physical parameters and processes at extremely low and zero metallicity that will likely be revealed with JWST and ground-based extremely large telescopes. 
One of the key observables is that of He {\sc ii} emission lines that might either originate from ionisation or stellar winds. Locally, VMS produce fast winds and broad He {\sc ii} emission lines (e.g. Crowther et al. 2016) while at lower $Z$ wind velocities should drop and lead to narrow {\it stellar} He {\sc ii} lines (Gr\"afener \& Vink 2015). Narrow lines might alternatively be entirely {\it nebular} which means we need to identify the ionising sources in different host galaxies. Again VMS might (e.g. Leitherer 2020) contribute, but stripped He stars could be a viable alternative (e.g. G\"otberg et al. 2020; Eldridge \& Stanway 2022.). Either way, the ionising radiation from He stars is a strong function of the wind optical depth, and thus requires a thorough understanding of hot-star winds across the full upper HRD.





\section*{DISCLOSURE STATEMENT}
The authors are not aware of any affiliations, memberships, funding, or financial holdings that
might be perceived as affecting the objectivity of this review. 

\section*{ACKNOWLEDGMENTS}
I would like to thank the Armagh Mdot gang: Erin Higgins, Gautham Sabhahit, and Andreas Sander for providing detailed comments on an earlier version of the manuscript. I also wish to acknowledge comments received from Jose Groh, Norbert Langer and Jo Puls. I extend my gratitude to many friends \& colleagues from the field who helped me in my understanding, especially my former students, postdocs, VFTS members, as well as Stan Owocki, Jon Sundqvist, John Hillier, Paula Benaglia, Nathan Smith, Jonathan Mackey, Luca Grassitelli, Raphael Hirschi, and others on specific issues related to this manuscript. Finally, I like to finish with a quote inspired by Acknowledgements of the late Adi Pauldrach in his P Cygni paper (with Jo Puls in 1990) that inspired my early PhD work on bi-stability - for which I thank my supervisors Alex de Koter and H. etc. Lamers. I thank God, Einstein, and the Big Bang for making massive stars available.

%


\noindent


\newpage


\section{Appendix: Mass-loss Recipes}
\label{sec_rec}

An oft-used combination of mass-loss recipes for massive stars is the {\sc dutch} Recipe in {\sc MESA}. It consists of a theoretical 
Vink et al. (2000, 2001) for hot OB stars, an empirical recipe of de Jager et al. (1988) for cooler supergiants below 10\,kK, and an empirical recipe by Nugis \& Lamers (2000) for 
WR stars, with a simple definition that $Y$ is larger than 0.4 and \teff\ is above 10\,kK.

For OB stars the most-oft used mass-loss recipe is that from Monte Carlo models of Vink et al. (2000) on either side of the BS-Jump.

\begin{eqnarray}
{\rm log}~\dot{M} & = &~-~6.697~(\pm 0.061) \nonumber \\
                  & &~+~2.194~(\pm 0.021)~{\rm log}(L_*/{10^5}) \nonumber \\
                  & &~-~1.313~(\pm 0.046)~{\rm log}(M_*/30) \nonumber\\
                  & &~-~1.226~(\pm 0.037)~{\rm log}\left(\frac{v_{\infty}/v_{\rm esc}}{2.0}\right) \nonumber \\
                  & &~+~0.933~(\pm 0.064)~{\rm log}(\teff/40 000) \nonumber\\
                  & &~-~10.92~(\pm 0.90)~\{{\rm log}(\teff/40 000)\}^{2} \nonumber\\
                  \nonumber\\
                  & &~{\rm for}~27~500 < \teff \le 50~000 {\rm K}
\label{eq_Ofit}
\end{eqnarray}
where $\dot{M}$ is in $\msun$ ${\rm yr}^{-1}$, $L_*$ and $M_*$ are in solar units
and $\teff$ is in Kelvin. 
In this range $v_{\infty}/v_{\rm esc}$ = 2.6. 
Equation ~\ref{eq_Ofit} predicts the calculated mass-loss rates of the 
180 models with a root-mean-square accuracy of 0.061 dex.
The second range (roughly the range of the B-type supergiants) is taken from
$\teff$ models below between 22~500 K:

\begin{eqnarray}
{\rm log}~\dot{M} & = &~-~6.688~(\pm 0.080) \nonumber \\
                  & &~+~2.210~(\pm 0.031)~{\rm log}(L_*/{10^5}) \nonumber \\
                  & &~-~1.339~(\pm 0.068)~{\rm log}(M_*/30) \nonumber\\
                  & &~-~1.601~(\pm 0.055)~{\rm log}\left(\frac{v_{\infty}/v_{\rm esc}}{2.0}\right) \nonumber \\
                  & &~+~1.07~(\pm 0.10)~{\rm log}(\teff/20 000) \nonumber\\
                  \nonumber\\
                  & &~{\rm for}~12~500 < \teff \le 22~500 {\rm K}
\label{eq_Bfit}                  
\end{eqnarray}
where again $\dot{M}$ is in $\msun$ ${\rm yr}^{-1}$, $L_*$ and $M_*$ 
are in solar units and $\teff$ is in Kelvin. In this range $v_{\infty}/v_{\rm esc}$ = 1.3. 
Vink et al. (2001) added a $Z$-dependence to this recipe and as Fe sets the mass-loss rate, this is mostly Fe-dependent mass loss.
Using locally consistent mass-loss rates, Vink \& Sander (2021) derived a shallower $\dot{M}$ - $Z$ dependence than in the original 
Vink et al. (2001) recipe.
Renzo et al. (2017) added a unitless $\eta$ parameter -- of order unity -- to study the effects of weaker winds in a similar vein to the Reimers mass-loss law for cool stars.

Bj\"orklund et al. (2021) constructed a modified wind momentum relation based on CMF computations as:
        \begin{equation}
          \log\left(\modf\right)  =  -1.55 + 0.46\log\left(\frac{Z_{\ast}}{Z_{\odot}}\right) + \\
                      \left[2.07 - 0.73\log\left(\frac{Z_{\ast}}{Z_{\odot}}\right)\right]\log\left(\frac{L_{\ast}}{10^{6}L_{\odot}}\right) 
        \end{equation}
        and
        \begin{equation}
          \log(\Mdot)  =  -5.55 + 0.79\log\left(\frac{Z_{\ast}}{Z_{\odot}}\right) + \\
                      \left[2.16 - 0.32\log\left(\frac{Z_{\ast}}{Z_{\odot}}\right)\right]\log\left(\frac{L_{\ast}}{10^{6}L_{\odot}}\right),
        \end{equation}
        with the wind momentum rate in units of $M_{\odot}$ yr$^{-1}$ km s$^{-1}$ $R_{\odot}^{0.5}$ and the mass-loss rate in units of $M_{\odot}$ yr$^{-1}$.
        When the complete model-grid sample is considered, these fitted relations give mean values that agree with the original simulations to within 10\%.
Note that there is no \teff\ dependence in this mass-loss recipe.

On the empirical side, de Jager et al. (1988) collected mass-loss rates determined for 271 Galactic stars of spectral types O through M. 
The mass-loss rates could be reproduced by a single formula dependent on \Teff\ and $L$:
\begin{equation}
\log \Mdot = -8.158 + 1.769 \log L - 1.676 \log \Teff\,.
\end{equation}
Here the \teff\ dependence is opposite to that in the Vink et al. (2000) recipe, which means it is not meaningful to compare these recipes in 
an $\dot{M}$-$L$ plot without accounting for the multi-variate -- and in this case even oppositely directed with regards to \teff\ -- behaviour of $\dot{M}$. Note that  
the original de Jager formulation did not account for stellar mass, and an improved relation was proposed by Nieuwenhuizen et al. (1990).

For WR stars the oft-used empirical relation of Nugis \& Lamers (2000) is:

\begin{eqnarray}
{\rm log}~\dot{M} & = &~-~11.0 ~+~1.29~(\pm 0.14)~{\rm log}L \nonumber \\
                  & &~+~1.73~(\pm 0.42)~{\rm log}Y ~+~0.47~(\pm 0.09)~{\rm log}Z 
                  \label{eq_NL}
\end{eqnarray}

For both WN and WC stars, and with with a standard
deviation of 0.19 dex. 
Note that this $Z$ dependence depends on the total self-enriched $Z$ and represents a very different $Z$ dependence from the Fe-dependent mass-loss rate of Vink \& de Koter (2005) and 
Sander \& Vink (2020):

\begin{eqnarray}
  \log \dot{M} =  a \cdot
\log\left[-\log\left(1-\Gamma_{\rm e}\right)\right] - \log(2) \cdot
\left(\frac{\Gamma_{\rm e,b}}{\Gamma_{\rm e}}\right)^c
+ \log \dot{M}_{\rm off}
\end{eqnarray}

with coefficients
\begin{eqnarray}
a = 2.932 (\pm 0.016)\\
    \Gamma_{\rm e,b}  = -0.324 (\pm 0.011) \cdot
\log(Z/Z_\odot) + 0.244 (\pm 0.010) \\
                   c = -0.44 (\pm 1.09) \cdot
\log(Z/Z_\odot) + 9.15 (\pm 0.96) \\
       \log \dot{M}_{\rm off} = 0.23 (\pm 0.04) \cdot \log(Z/Z_\odot)
- 2.61 (\pm 0.03) 
\end{eqnarray}


For stripped He stars with masses and luminosities below the classical WR regime the only predictions available are the Monte Carlo computations by Vink (2017):

\begin{equation}
\label{eq_He}
\log \mdot~=~-13.3~+~1.36 \log(L/\lsun)~+~0.61 \log(Z/\zsun)    
\end{equation}
with a fitting error of $\sigma$ = 0.11. The application of this formula is only recommended for lower-mass stripped stars, and not for
classical WR stars. This stripped-star formula could be used as a lower bound to a classical WR recipe such as that of Sander \& Vink (2020).

\end{document}